\documentclass[%
reprint,
groupedaddress,
nofootinbib,
 amsmath,amssymb,
 aps,
pra,
floatfix,
longbibliography
]{revtex4-1}

\usepackage{graphicx}
\usepackage{dcolumn}
\usepackage{bm}
\usepackage{braket}
\usepackage{physics}
\usepackage{comment}
\usepackage{color}
\usepackage[version=3]{mhchem}
\usepackage{ulem}

\usepackage[whole]{bxcjkjatype} 
\usepackage[colorlinks=true,urlcolor=blue,citecolor=blue,linkcolor=blue,breaklinks=true]{hyperref}

\begin{document}

\preprint{APS/123-QED}

\title{Interaction induced topological magnon in electron-magnon coupled systems}

\author{Kosuke Fujiwara}
\affiliation{Department of Applied Physics, The University of Tokyo, Hongo, Tokyo, 113-8656, Japan}

\author{Takahiro~Morimoto}
\affiliation{Department of Applied Physics, The University of Tokyo, Hongo, Tokyo, 113-8656, Japan}

\date{\today}

\begin{abstract}
We theoretically study the emergence of topological magnons in electron-magnon coupled systems. 
The magnon dispersion in a ferromagnet usually possesses an effective time reversal symmetry in the absence of Dzyaloshinskii-Moriya (DM) interaction, preventing the appearance of topological magnons.
When a spin system is coupled to itinerant electrons, we find that the magnon band structure of the spin system experiences time-reversal symmetry breaking with the electron-magnon interaction via the exchange coupling, where topological magnons arise without requiring strong DM.
Specifically, we consider a heterostructure consisting of a ferromagnetic insulator and a transition metal dichalcogenide (TMD) monolayer and investigate topological gap opening in magnon bands. 
Our findings reveal that even trivial ferromagnets can host topological magnons via coupling to itinerant electronic systems.
\end{abstract}

\maketitle



\textit{Introduction.}---
Topology and geometry play a central role in studies on quantum materials~\cite{Hasan2010Colloquium:Insulators,Qi2011TopologicalSuperconductors}. Intrinsic topological properties of electrons, such as the quantum Hall effect~\cite{Klitzing1980NewResistance,Thouless1982QuantizedPotential}, the quantum anomalous Hall effect~\cite{Haldane1988ModelAnomaly,Onoda2003QuantizedMetals}, and the quantum spin Hall effect~\cite{Murakami2003DissipationlessTemperature,Kane2005QuantumGraphene,Bernevig2006QuantumWells}, are attributed to nontrivial quantum geometry exemplified by the Chern number and the Berry curvature in momentum space~\cite{Xiao2010BerryProperties}.
While the topological phases of electrons are usually discussed for their ground state properties, the excited states of many-body quantum systems also support rich topological structures. As a representative example, topological properties of bosonic excitations in magnetic materials, magnons, have attracted significant attention~\cite{Katsura2010TheoryMagnets,Onose2010ObservationEffect,Ideue2012EffectInsulators,Hirschberger2015ThermalMagnet,Zhang2010TopologicalEffect}.
In particular, topological properties of magnons have been intensively studied because they provide promising candidates for Joule-heat-free information transmission and processing~\cite{Chumak2015MagnonSpintronics}. The Berry curvature and Chern number of magnons lead to various intriguing phenomena, such as the thermal Hall effect~\cite{Matsumoto2011RotationalEffect,Matsumoto2011TheoreticalFerromagnets,Matsumoto2014ThermalInteraction}, the spin Nernst effect~\cite{Cheng2016SpinAntiferromagnets,Zyuzin2016MagnonAntiferromagnets}, and the nonlinear magnon spin Nernst effect~\cite{Kondo2022NonlinearCurrent}.

Realization of topological magnons with nonzero Chern numbers necessitates the breaking of an effective time-reversal symmetry (TRS) of the spin system arising from a combination of time-reversal operation and spin rotation~\cite{Mook2019ThermalAntiferromagnets}.
In previous studies, the Dzyaloshinskii-Moriya (DM) interaction~\cite{Owerre2016AInsulator,Owerre2017TopologicalAntiferromagnets,Laurell2018MagnonInteractions,Mook2019ThermalAntiferromagnets,Kawano2019ThermalAntiferromagnet} or geometrically nontrivial spin textures~\cite{Owerre2017MagnonInteraction,Rosales2019FromInsulators,Kim2019MagnonAntiferromagnet,GomezAlbarracin2021ChiralLattice,Fujiwara2022ThermalSystems} are considered to break the TRS. The DM interaction typically arises from spin-orbit coupling (SOC) in magnetic materials, which requires heavy elements to enhance the SOC strength. More recently, interaction-induced topological magnons have been investigated. For example, theoretical studies suggest that topological magnons can emerge from magnon-magnon interactions originating from the DM interaction~\cite{Mook2021Interaction-StabilizedFerromagnets,Chatzichrysafis2025ThermalScattering}, or through magnon-phonon interactions~\cite{Go2019TopologicalNumbers,Park2019TopologicalAntiferromagnets,Park2020ThermalInteraction}.

Electron-magnon interactions provide a novel pathway to realize topological magnons. While the coupling between electrons and magnons has been extensively investigated, previous studies have primarily focused on its effects on the electronic system, such as magnon-mediated superconductivity~\cite{Kargarian2016AmpereanInsulators,Rohling2018SuperconductivityMagnons,Mland2024Many-bodyAltermagnets}, topological superconductivity~\cite{Mland2023TopologicalMagnons,Mland2023TopologicalStates,VinasBostrom2024Magnon-mediatedWire}, and the realization of topological electronic phases~\cite{Fujiwara2025TopologicalInteractions}. In contrast, the modification of the magnon topology by the itinerant electrons has remained largely unexplored.

\begin{figure}[tb]
    \centering
    \includegraphics[width=\linewidth]{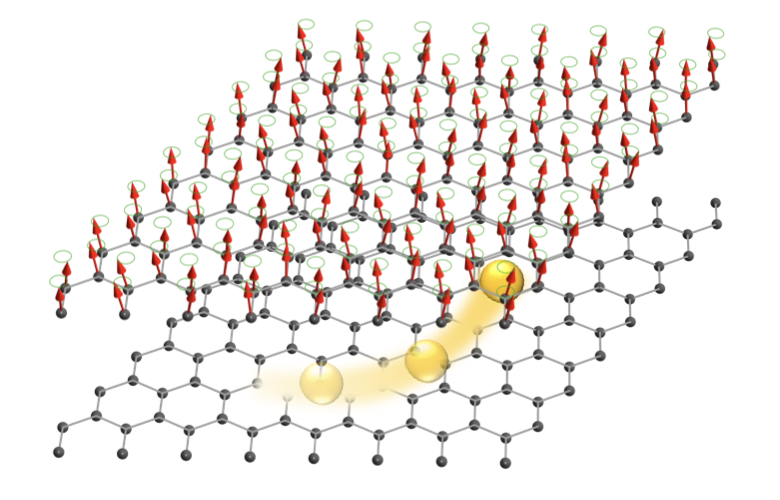}
    \caption{Schematic of a spin system coupled to an electronic system. Yellow balls indicate a flow of electrons. Geometrical properties of the electron system with broken time-reversal symmetry are imprinted on the magnon system via electron-magnon interactions, resulting in the emergence of topological magnons.} 
    \label{fig:schematic}
\end{figure}

In this paper, we demonstrate that magnons can acquire nontrivial topology via electron-magnon interactions in a heterostructure consisting of a ferromagnetic insulator and a two-dimensional electron system (see Fig.~\ref{fig:schematic}). Through the proximity effect, an effective magnetic field is induced in the electron system, explicitly breaking its TRS. This symmetry breaking is then imprinted onto the magnon system through electron-magnon interactions, resulting in the emergence of topological magnons. Specifically, we analyze a model composed of a transition metal dichalcogenide (TMD) monolayer coupled to a ferromagnet. We show that a sizable topological gap opens in the magnon spectrum, which can be controlled by tuning the chemical potential of the electron system.

\textit{Model.}---
We consider a system composed of a ferromagnetic insulator on a honeycomb lattice coupled to a transition metal dichalcogenide (TMD) monolayer. The localized spins interact with itinerant electrons via the exchange interaction. The total Hamiltonian is given by
\begin{equation}
    \mathcal{H}_{\text{tot}} = \mathcal{H}_{\text{s}} + \mathcal{H}_{\text{e}} + \mathcal{H}_{\text{int}},
\end{equation}
where $\mathcal{H}_{\text{s}}$ is the spin Hamiltonian, $\mathcal{H}_{\text{e}}$ is the electron Hamiltonian, and $\mathcal{H}_{\text{int}}$ is the interaction between itinerant electrons and localized spins. The spin Hamiltonian is given by the ferromagnetic Heisenberg model
\begin{equation}
    \mathcal{H}_{\text{s}} = -J\sum_{\langle i,j\rangle}\vb*{S}_i\cdot \vb*{S}_j-\Delta_z\sum_i(S_i^{z})^2,\label{eq:spin_hamiltonian}
\end{equation}
where $\vb*{S}_i$ is the spin operator at site $i$, $J$ is the exchange coupling, and $\Delta_z$ is the easy-axis anisotropy. We assume a ferromagnetic ground state with spins aligned along the $z$ axis.

The TMD monolayer forms a honeycomb lattice, and we employ a minimal two-band tight-binding model to capture its band structure around $K$ and $K^\prime$ points, showing massive Dirac cones \cite{Liu2013Three-bandDichalcogenides}. The two-band tight-binding model consists of a honeycomb lattice with a large staggered potential. To incorporate the effect of SOC, we employ a spin-dependent complex hopping between next-nearest neighbors used in the Haldane model~\cite{Haldane1988ModelAnomaly, Ezawa2012Valley-PolarizedSilicene}.
Specifically, the Hamiltonian of TMDs is given by
\begin{align}
    \mathcal{H}_{\text{e}}=&-t_1\sum_{\langle i,j\rangle}\sum_{\mu}c^\dagger_{i,\mu} c_{j,\mu} - t_2\sum_{\langle\langle i,j\rangle\rangle}\sum_{\mu}e^{i\chi_{ij}\phi_{\mu}}c_{i,\mu}^\dagger c_{j,\mu} &\notag\\
    &+ \Delta\sum_i\sum_{\mu} \xi_i c_{i,\mu}^\dagger c_{i,\mu},
\end{align}
where $c^\dagger_{i,\mu}$ is a creation operator of an electron at site $i$ with spin $\mu(=\uparrow,\downarrow)$, $t_1$ and $t_2$ are the nearest and next-nearest-neighbor hopping amplitudes, respectively.
The phase factor $\phi_{\mu}$ depends on the spin $\mu$ as $\phi_{\uparrow}=\phi$ and $\phi_{\downarrow}=-\phi$, and $\chi_{ij}=+1(-1)$ for the next-nearest-neighbor hopping in the counterclockwise direction for the $A$ ($B$) sublattice. $\Delta$ is the staggered potential, and $\xi_i=+1(-1)$ for sublattice $A(B)$.

\begin{figure}[tb]
    \centering
    \includegraphics[width=\linewidth]{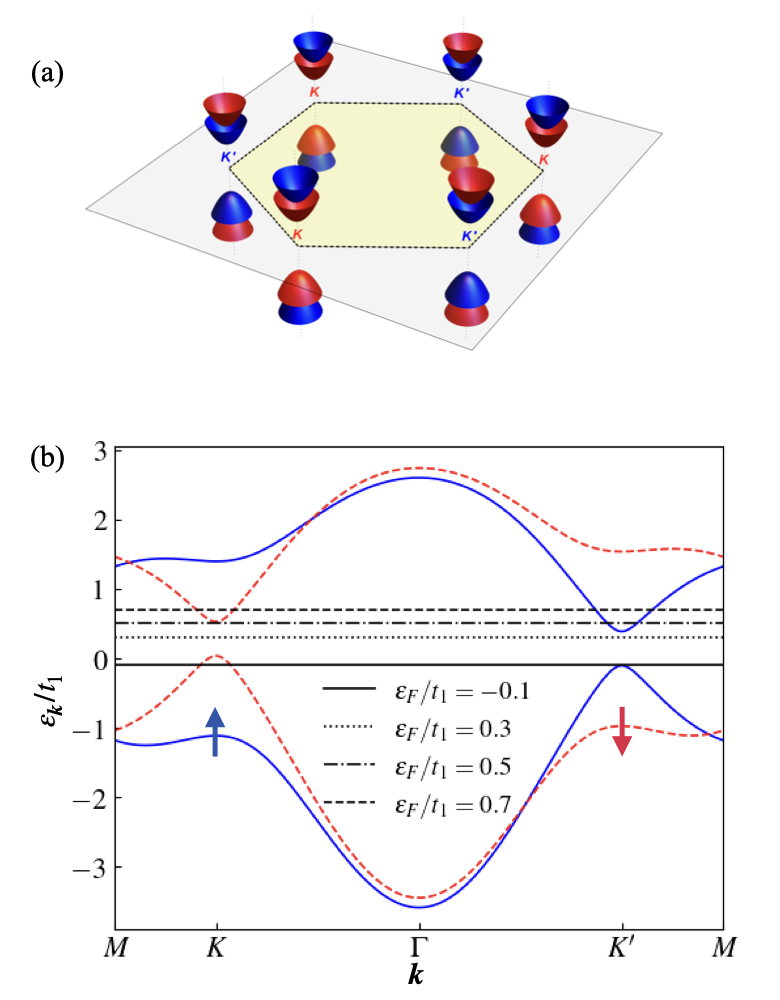}
    \caption{Schematic of the TMD monolayer and its band structure. (a) Schematic of the TMD monolayer. The band structure is gapped due to the large staggered potential $\Delta$. Blue and red cones denote spin-up and spin-down bands, respectively. (b) Band structure of the TMD monolayer. Blue solid and red dashed lines denote spin-up and spin-down bands, respectively. Vertical lines indicate the position of the Fermi energy $\varepsilon_F$. We use the following parameters: $t_2/t_1=0.12$, $\Delta/t_1=0.75$, $\phi=0.3\pi$, $J/t_1=1.0\times10^{-3}$, $\Delta_z/t_1=1.0\times10^{-4}$, $J_{ex}/t_1=0.07$, and $S=1$.} 
    \label{fig:TMD_ele}
\end{figure}

Inversion symmetry is broken in the TMD monolayer due to the staggered potential $\Delta$, whereas TRS is preserved since the phase factor $\phi$ has opposite signs for spin-up and spin-down electrons. 
The band structure of TMD is gapped due to the large staggered potential $\Delta$. Typical model parameters are given by $t_1\sim1$ eV, $t_2\sim100$ meV, and $\Delta\sim750$ meV~\cite{Xiao2012CoupledDichalcogenides}. Thus, the Chern number of each band is usually zero because the condition $\Delta > 3\sqrt{3} t_2\sin{\phi}$ holds. A schematic of TMD is shown in Fig.~\ref{fig:TMD_ele} (a).

The interaction between electrons and spins is given by
\begin{equation}
    \mathcal{H}_{\text{int}} = -J_{ex}\sum_{i,\mu,\nu}\vb*{S}_{i}\cdot c_{i\mu}^\dagger\vb*{\sigma}_{\mu,\nu}c_{i\nu},\label{eq:interaction}
\end{equation}
where $\vb*{\sigma}$ are the Pauli matrices, and $J_{ex}$ is the exchange coupling. 
The exchange coupling $J_{ex}$ typically ranges from a few to a few tens of meV~\cite{Zhong2017VanValleytronics,Zhao2017EnhancedField,Norden2019GiantEffect,Zollner2019ProximityDependence}, while it can be a few hundred meV under high pressure~\cite{Zhang2018StrongHeterostructures,Li2022GiantFerromagnetism}. 
By applying the Holstein-Primakoff transformation in the linear spin-wave approximation~\cite{Holstein1940FieldFerromagnet}
\begin{equation}
    S_{j}^+ \approx \sqrt{2S}a_j,~
    S_{j}^- \approx \sqrt{2S}a_j^\dagger,~
    S_{j}^z = S-a_j^\dagger a_j,
\end{equation}
to the spin Hamiltonian $\mathcal{H}_{\text{s}}$ and the interaction $\mathcal{H}_{\text{int}}$, we obtain the magnon Hamiltonian $\mathcal{H}_{\text{m}}$ and electron-magnon interactions. Here, $a_j$ is the annihilation operator of the magnon at the $j$-th site. 
The interaction in Eq.\eqref{eq:interaction} contains the static magnetic exchange field acting on the electron system. We define $\mathcal{H}_{\text{e}}^\prime$ as the sum of the electron Hamiltonian $\mathcal{H}_{\text{e}}$ and the effective magnetic field induced by the interaction Eq.\eqref{eq:interaction} as
\begin{equation}
    \mathcal{H}_{\text{e}}^\prime = \mathcal{H}_{\text{e}}-J_{ex}S\sum_i (c_{i,\uparrow}^\dagger c_{i,\uparrow}-c_{i,\downarrow}^\dagger c_{i,\downarrow}).
\end{equation}
This exchange field breaks TRS of the electron system. The band structure of the electron Hamiltonian $\mathcal{H}_{\text{e}}^\prime$ is shown in Fig.~\ref{fig:TMD_ele} (b). The field acts as a Zeeman term, splitting the energies of the spin-up and spin-down bands. When the Fermi energy $\varepsilon_F$ crosses $K$ or $K^\prime$ valley with specific spin, the electron system breaks TRS effectively since the integrated Berry curvature of occupied bands becomes non-zero.
Then, the total Hamiltonian can be approximated and rewritten as 
$\mathcal{H}_{\text{tot}} \approx  \mathcal{H}_{\text{m}} + \mathcal{H}_{\text{e}}^\prime + \mathcal{H}_{\text{int}_1} +  \mathcal{H}_{\text{int}_2}$ with 
$\mathcal{H}_{\text{m}}=\sum_{\vb*{k}}\psi^\dagger_{\text{m},\vb*{k}}\omega_{\vb*{k}}\psi_{\text{m},\vb*{k}}$,
$\mathcal{H}_{\text{e}}^\prime=\sum_{\vb*{k}}\sum_\mu\psi_{\text{e},\vb*{k},\mu}^\dagger \varepsilon_{\vb*{k},\mu}\psi_{\text{e},\vb*{k},\mu}$ where $\psi_{\text{m},\vb*{k}}$ and $\psi_{\text{e},\vb*{k},\mu}$ are the magnon and electron operators in the band basis, respectively. The matrices $\omega_{\vb*{k}}$ and $\varepsilon_{\vb*{k},\mu}$ correspond to the diagonal matrices of eigenvalues for the magnon Hamiltonian $\mathcal{H}_{\text{m}}$ and the electron Hamiltonian $\mathcal{H}_{\text{e}}^\prime$, respectively. The interaction terms are given as 
\begin{subequations}
    \begin{align}
        \mathcal{H}_{\text{int}_1}=&\sum_{\vb*{k},\vb*{q}}\sum_{\alpha,\beta,\gamma}\frac{V_{\vb*{k},\vb*{q}}^{\alpha\beta\gamma}}{\sqrt{N}}\psi^\dagger_{\text{e},\vb*{k},\downarrow,\alpha}\psi_{\text{e},\vb*{q},\uparrow,\beta}\psi_{\text{m},\vb*{k}-\vb*{q},\gamma}+h.c.\\ \mathcal{H}_{\text{int}_2}=&\sum_{\vb*{k},\vb*{q},\vb*{p}}\sum_{\mu}\sum_{\alpha,\beta,\gamma,\delta}\frac{W_{\vb*{k},\vb*{q},\vb*{p}}^{\mu,\alpha\beta\gamma\delta}}{N} \nonumber \\
        & \times 
        \psi^\dagger_{\text{e},\vb*{k},\mu,\alpha}\psi_{\text{e},\vb*{q},\mu,\beta}\psi^\dagger_{\text{m},\vb*{p},\gamma}\psi_{\text{m},\vb*{k}-\vb*{q}+\vb*{p},\delta},
    \end{align}
\end{subequations}
where $N$ is the number of unit cells, $\alpha$, $\beta$, $\gamma$, and $\delta$ are the band indices.
Here, $\mathcal{H}_{\text{int}_1}$ and
$\mathcal{H}_{\text{int}_2}$ represent interactions between electrons and magnons that change and conserve the number of magnons, respectively. $V$ and $W$ are the interaction coefficients in the band basis. The details of the Hamiltonian are described in Appendix \ref{sec:appendix_model}.

\textit{Effective Hamiltonian and topological magnon.}---
Now, we derive an effective Hamiltonian using the Schrieffer-Wolff transformation.
The effective Hamiltonian is given by
\begin{align}
    H_{\text{eff},\vb*{k}}^{\alpha\beta} =& \omega_{\vb*{k}}^{\alpha\beta}+\frac{1}{N}\sum_{\vb*{q}}\sum_{\mu}\sum_{\gamma}W^{\mu,\gamma\gamma\alpha\beta}f_F(\varepsilon_{\vb*{q},\mu,\gamma})\notag\\
    &-\frac{1}{2N}\sum_{\vb*{q}}\sum_{\gamma,\delta} V_{\vb*{q},\vb*{q}-\vb*{k}}^{\gamma\delta\beta}(V_{\vb*{q},\vb*{q}-\vb*{k}}^{\gamma\delta\alpha})^*\notag\\
    &\times\frac{1}{2}\bigg[\frac{(f_F(\varepsilon_{\vb*{q},\gamma})-f_F(\varepsilon_{\vb*{q}-\vb*{k},\delta}))}{\omega_{\vb*{k},\beta}-\varepsilon_{\vb*{q},\downarrow,\gamma}+\varepsilon_{\vb*{q}-\vb*{k},\uparrow,\delta}}+(\alpha\leftrightarrow\beta)\bigg].\label{eq:Heff}
\end{align}
The details of the derivation are in Appendix~\ref{sec:appendix_effective_hamiltonian}.
This approximation is valid when the magnon energy $\omega_{\vb*{k},\beta}$ does not overlap with the particle-hole excitation continuum of electrons $\varepsilon_{\vb*{q},\downarrow,\gamma}-\varepsilon_{\vb*{q}-\vb*{k},\uparrow,\delta}$ where $f_F(\varepsilon_{\vb*{q},\gamma})-f_F(\varepsilon_{\vb*{q}-\vb*{k},\delta})\neq0$. This condition is satisfied when the Fermi energy lies within the band gap of the electron system. Since the magnon energy is smaller than the electronic band gap, the magnon band does not overlap with the particle-hole continuum of electrons in this case. 
Even when the electron system is in a metallic state (i.e., the Fermi energy lies within the band), this condition can be satisfied provided that the Fermi surface is small or located in a specific valley. In such cases, kinematic constraints prevent the particle-hole continuum from overlapping with the magnon energy.

\begin{figure}[tb]
    \centering
    \includegraphics[width=\linewidth]{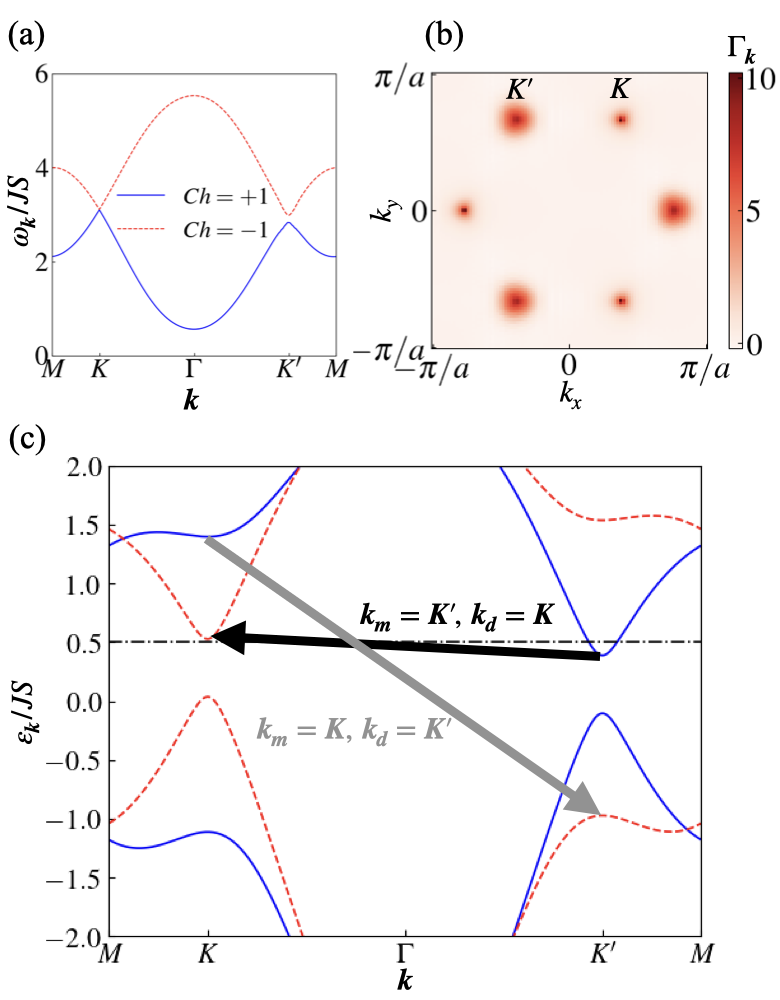}
    \caption{Band structure and  Berry curvature of the effective Hamiltonian coupled to TMDs. (a) Band structure and Chern number of the effective Hamiltonian. (b) Berry curvature of the lowest band in log-scale $\Gamma(\Omega)=\text{sign}(\Omega)\log(1+|\Omega|)$. 
    (c) Band structure of the TMD monolayer around the Fermi level. Blue solid and red dashed lines represent spin-up and spin-down bands, respectively. Black and gray arrows depict transitions that make dominant contributions to magnons at the $K^\prime$ and $K$ points, respectively. $\vb{k_m}$ and $\vb{k_d}$ denote the momenta of the magnon and the spin-down electron, respectively, which correspond to $\vb*{k}$ and $\vb*{q}$ in Eq.~\eqref{eq:Heff}. We use the following parameters: $t_2/t_1=0.12$, $\Delta/t_1=0.75$, $\phi=0.3\pi$, $J/t_1=1.0\times10^{-3}$, $\Delta_z/t_1=1.0\times10^{-4}$, $J_{ex}/t_1=0.07$, $S=1$ and $\varepsilon_F/t_1=0.5$}
    \label{fig:TMD_magnon}
\end{figure}

Using this effective Hamiltonian, we can study the magnon band structure renormalized by the electron-magnon interaction. The magnon band acquires a gap at the $K$ and $K^\prime$ points through the electron-magnon interaction as shown in Fig.~\ref{fig:TMD_magnon} (a). In our model, the magnon Hamiltonian itself is gapless at the $K$ and $K^\prime$ points. However, since the TMDs with magnetic exchange interaction break the TRS and the inversion symmetry, a nontrivial geometry of the electron system is imprinted onto the magnon system via electron-magnon interactions. Then, a topological gap opens at the $K$ and $K^\prime$ points.
Furthermore, the magnon band has a nontrivial Chern number as shown in Fig.~\ref{fig:TMD_magnon} (a). To validate this effective Hamiltonian approach, we calculated the magnon spectral function using the Green's function method. We confirmed that the band structure of the effective Hamiltonian agrees well with the spectral function and that the magnons have a long lifetime (see Appendix~\ref{sec:appendix_spectral} for details).
Figure~\ref{fig:TMD_magnon}(b) displays the Berry curvature of the lowest band. The Berry curvature exhibits peaks at the $K$ and $K^\prime$ points with the same sign, resulting in a non-zero Chern number for the lowest band.
In particular, for magnons with momenta around the $K$ ($K^\prime$) point, the third term in Eq.~\eqref{eq:Heff} is dominated by contributions from spin-down electrons around the $K^\prime$ ($K$) point and spin-up electrons around the $K$ ($K^\prime$) point due to the small energy difference $\varepsilon_{\vb*{q},\downarrow,\gamma}-\varepsilon_{\vb*{q}-\vb*{k},\uparrow,\delta}$. We show schematics of these transitions in Fig.~\ref{fig:TMD_magnon} (c) when the Fermi energy intersects the spin-up band at the $K^\prime$ valley ($\varepsilon_F/t_1=0.5$). The dominant transitions for magnons at $k=K^\prime$ and $k=K$ are indicated by the black and gray arrows, respectively. Since the energy difference corresponding to the black arrow is smaller than that of the gray arrow, a larger topological gap opens at the $K^\prime$ point.

\begin{figure}[tb]
    \centering
    \includegraphics[width=\linewidth]{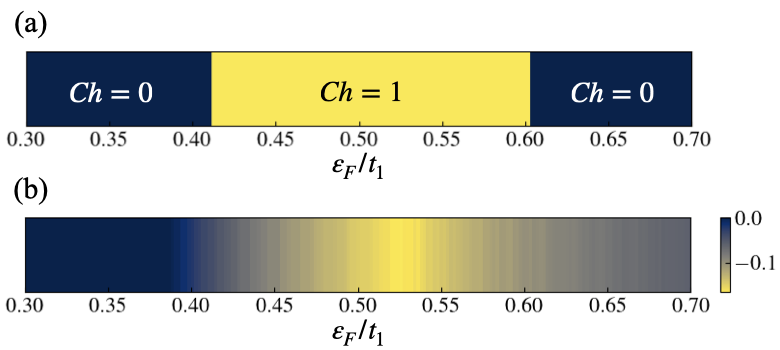}
    \caption{Topological phase diagram with the Fermi energy dependence. (a) Chern number of the lowest magnon band as a function of the Fermi energy. (b) Integrated Berry curvature of occupied bands of the electron system as a function of the Fermi energy. We use the following parameters: $t_2/t_1=0.12$, $\Delta/t_1=0.75$, $\phi=0.3\pi$, $J/t_1=1.0\times10^{-3}$, $\Delta_z/t_1=1.0\times10^{-4}$, $J_{ex}/t_1=0.07$, and $S=1$.} 
    \label{fig:TMD_phase}
\end{figure}

Next, we show the Fermi-energy dependence of the Chern number of the lowest magnon band in Fig.~\ref{fig:TMD_phase} (a). When the Fermi energy lies within the band gap of TMDs ($\varepsilon_F/t_1=0.3$), the Chern number of the lowest magnon band is zero, in contrast to the above metallic case with a nonzero Chern number at $\varepsilon_F/t_1=0.5$. This is because the Berry curvature of the occupied electron bands has peaks at the $K$ and $K^\prime$ points with opposite signs. Consequently, the integrated Berry curvature over the occupied bands becomes finite when the Fermi energy intersects only one valley, where the topological properties of the electron system are imprinted onto the magnon system via electron-magnon interactions. In Fig.~\ref{fig:TMD_phase}(b), we show the integrated Berry curvature of the occupied electron bands. We find that the energy level where the magnon Chern number is non-zero corresponds to the energy range where the electron system exhibits a large integrated Berry curvature.

Finally, we comment on the valley selectivity of this effect. 
In this model, topological magnons emerge when the Fermi energy intersects the spin-up band at the $K^\prime$ valley ($\varepsilon_F/t_1=0.5$ as shown in Fig.~\ref{fig:TMD_ele} (b)). By tuning the chemical potential, the Fermi energy can also intersect the spin-down band at the $K$ valley as shown for $\varepsilon_F/t_1=-0.1$ in Fig.~\ref{fig:TMD_ele} (b). In this case, the magnon around the $K^\prime$ point also strongly interacts with the electrons, and a large gap opens at the $K^\prime$ point. Notably, the Chern number of the lowest magnon band remains zero in this case since the Berry curvature of the magnon band has opposite signs at the $K$ and $K^\prime$ points. 
This is because the transition processes contributing to the effective magnon Hamiltonian around the $K^\prime$ points are different for $\varepsilon_F/t_1=-0.1$, when compared to the case with $\varepsilon_F/t_1=0.5$ as detailed in Appendix~\ref{sec:appendix_valley}. 
Specifically, the sign of the Berry curvature around the $K^\prime$ points becomes opposite to the case with $\varepsilon_F/t_1=0.5$, resulting in zero Chern number for the lowest magnon band.

\textit{Discussion.}---
In this study, we demonstrate the emergence of topological magnons induced by electron-magnon interactions. In particular, we focus on a spin system coupled to a TMD monolayer, where the topological magnon appears due to the electron-magnon interactions.

By using the typical parameters for TMDs and the exchange interaction, we confirm that a sizable topological gap can be induced in the magnon system. 
Using realistic parameters (\cite{Xiao2012CoupledDichalcogenides}) used for the magnon dispersion in Fig.~\ref{fig:TMD_magnon}(a), we estimate the gap size to be about $0.1J$ at the $K^\prime$ point. Candidate materials include heterostructures composed of TMD monolayers and 2D ferromagnets, such as the chromium trihalides \ce{CrX_3} ($X$=\ce{I}, \ce{Br}, \ce{Cl}) and \ce{CrXTe_3} ($X$ = \ce{Si}, \ce{Ge})~\cite{Gong2019Two-dimensionalDevices}. Among these, compounds consisting of lighter elements, such as \ce{CrBr_3}~\cite{Yelon1971RenormalizationEffects,Pershoguba2018DiracFerromagnets} and \ce{CrCl_3}~\cite{Schneeloch2022GaplessCrCl3,Chen2022MasslessCrCl3}, are particularly suitable for realizing the proposed mechanism. Here, we note that \ce{CrCl_3} requires an external magnetic field to induce out-of-plane ferromagnetic order~\cite{Lu2020Meron-likeCrCl3}. In these materials, the intrinsic spin-orbit coupling is weak, leading to a negligible DM interaction. Thus, topological features observed can be attributed to the electron-magnon interaction, as proposed in the present study. 
Experimentally, the topological magnon gap could be detected through inelastic neutron scattering experiments~\cite{Chen2018TopologicalCrI3,Zhang2020TopologicalMagnet}, and the edge states can be observed by inelastic electron tunneling spectroscopy~\cite{Zhang2026ObservationStates}.

While we have focused on ferromagnets coupled with TMDs, the proposed mechanism for topological magnons in this paper is generic and applicable to other systems with strong spin-orbit coupling, such as graphene with Rashba SOC. In heterostructures composed of graphene and a spin system, the Rashba-type SOC arises from broken mirror symmetry along the $z$ axis due to the stacking and the proximity coupling can induce magnetic exchange interaction, introducing TRS breaking to the electron system~\cite{Qiao2010QuantumEffects,Zhang2018StrongHeterostructures}. Thus, the electron system becomes a Chern insulator, and its topological properties are imprinted on the magnon system via electron-magnon interactions.

The results for graphene with Rashba SOC are presented in Appendix~\ref{sec:appendix_graphene}. Similar to the TMD case, the magnon band also becomes topological via electron-magnon interactions. Notably, the magnon band acquires nontrivial topology even when the staggered potential is large and the electron system is topologically trivial.

We now discuss the magnitude of the topological gap. The topological magnon phase proposed here originates from the second-order perturbation of the electron-magnon interaction. Consequently, the gap size is proportional to $J_{ex}^2/\varepsilon_{e}$. Here, $\varepsilon_{e}$ is the characteristic energy scale of the electron system, such as the band gap or the bandwidth.
Therefore, a system with a narrow bandwidth and strong exchange interaction $J_{ex}$ is preferable for realizing a large topological gap, where the typical exchange interaction $J_{ex}$ is $10\sim100$ meV. For graphene, we estimate the topological magnon gap to be of the order of $10^{-3}\times J$ (see Appendix~\ref{sec:appendix_graphene} for details). This value is smaller than that in TMDs due to the large bandwidth of graphene. However, even in graphene, the bandwidth can be reduced by forming a moiré lattice, such as in twisted bilayer graphene~\cite{Andrei2020GrapheneTwist}. Therefore, a larger gap can be anticipated in moiré heterostructures, such as twisted bilayer graphene coupled to a spin system, or a graphene/ferromagnetic-insulator bilayer with a relative twist angle.  In these systems, $J_{ex}$ is spatially dependent, making the interaction more complex. Nevertheless, a large topological gap is expected as long as the spatially-averaged coupling remains sufficiently large.

\acknowledgements
T.M. was supported by 
JSPS KAKENHI Grant 23K25816, 23K17665, 24K00568 and 24H02231.
K.F. was supported by JSPS KAKENHI Grant 24KJ0730 and the Forefront Physics and Mathematics program to drive transformation (FoPM).

\appendix

\section{Details of Model}\label{sec:appendix_model}
This appendix provides the explicit forms of the magnon Hamiltonian ($\mathcal{H}_{\text{m}}$), electron Hamiltonian ($\mathcal{H}_\text{e}$), and the interaction between electrons and magnons ($\mathcal{H}_{\text{int}1}$, $\mathcal{H}_{\text{int}2}$) in momentum space.
Since we assume that the spins are aligned along the $z$-direction, the magnon Hamiltonian is given by
\begin{align}
    \mathcal{H}_{\text{s}} \approx \mathcal{H}_{\text{m}} = \sum_{\vb*{k}}\phi^\dagger_{\text{m},\vb*{k}}H_{\text{m},\vb*{k}}\phi_{\text{m},\vb*{k}},
\end{align}
where $\phi_{\text{m},\vb*{k}}=(a_{\vb*{k}},b_{\vb*{k}})^T$, $a_{\vb*{k}} = \frac{1}{\sqrt{N}} \sum_{j \in A} e^{-i\vb*{k}\cdot\vb*{R}_j} a_j$ (and similarly for $b_{\vb*{k}}$ with $j \in B$) and $a_j$ ($b_j$) is the annihilation operator of the magnon at site $j$ on the A (B) sublattice. The Hamiltonian matrix of the magnon system $H_{\text{m},\vb*{k}}$ is given by 
\begin{equation}
    H_{\text{m},\vb*{k}} = 
    \begin{pmatrix}
         3JS+2S\Delta_z & -JS\gamma_{\vb*{k}}\\
        -JS\gamma_{\vb*{k}}^* & 3JS+2S\Delta_z
    \end{pmatrix}.
\end{equation}
Here, $\gamma_{\vb*{k}}=\sum_l e^{i\vb*{k}\cdot\vb*{\delta}_l}$, and $\vb*{\delta}_l$ are the nearest neighbor vectors $\vb*{\delta}_l=(0,-a),(\pm\sqrt{3}a/2,a/2)$.

The electron Hamiltonian in momentum space is given by
\begin{subequations}
\begin{align}
    \mathcal{H}_{\text{e}}&=\sum_{\vb*{k}}\sum_\mu\phi^\dagger_{\text{e},\vb*{k},\mu}H_{\text{TMD},\vb*{k},\mu}\phi_{\text{e},\vb*{k},\mu}\\
    \phi_{\text{e},\vb*{k},\mu}&=(c_{\vb*{k},\mu,A},c_{\vb*{k},\mu,B})^T,
\end{align}
\end{subequations}
where $c_{\vb*{k},\mu,l}$ is the annihilation operator of an electron with momentum $\vb*{k}$, spin $\mu(=\uparrow,\downarrow)$, and sublattice $l(=A,B)$. The explicit form of $H_{\text{TMD},\vb*{k},\mu}$ is given by
\begin{align}
    H_{\text{TMD},\vb*{k},\mu}=
    \begin{pmatrix}
     -t_2f(\phi_\mu,\vb*{k})+\Delta  & -t_1\gamma_{\vb*{k}} \\
     -t_1\gamma^*_{\vb*{k}}    & -t_2f(-\phi_\mu,\vb*{k})-\Delta 
    \end{pmatrix},
\end{align}
where $\phi_\mu$ is the spin-dependent phase factor defined by $\phi_{\uparrow}=\phi$ and $\phi_{\downarrow}=-\phi$ and the geometrical factor $f(\phi_\mu,\vb*{k})$ is defined as $f(\phi_\mu,\vb*{k})=2\cos(\phi_\mu)\sum_l\cos{(\vb*{k}\cdot\vb*{d}_l)}-2\sin(\phi_\mu)\sum_l\sin{(\vb*{k}\cdot\vb*{d}_l)}$. Here, $\vb*{d}_l$ denotes the next-nearest-neighbor vectors $\vb*{d}_l=(\sqrt{3}a,0),(\sqrt{3}a/2,3a/2),(-\sqrt{3}a/2,3a/2)$.

The interaction between electrons and magnons is given by
\begin{align}
    \mathcal{H}_{\text{int}} \approx & \mathcal{H}_{\text{int}_1} + \mathcal{H}_{\text{int}_2},\notag\\
=& -J_{ex}\sqrt{2S}\sum_{j}c^\dagger_{j,\downarrow}c_{j,\uparrow}a_{j}+h.c.\notag\\
&+ J_{ex}\sum_{j}(n_{j,\uparrow}-n_{j,\downarrow})a^\dagger_{j}a_j.
\end{align}
The interaction Hamiltonian in momentum space is written as
\begin{subequations}
\begin{align}
\mathcal{H}_{\text{int}_1}=& \sum_{\vb*{k},\vb*{q}}\sum_{l}\frac{\tilde{V}}{\sqrt{N}}\phi^\dagger_{\text{e},\vb*{k},\downarrow,l}\phi_{\text{e},\vb*{q},\uparrow,l}\phi_{\text{m},\vb*{k}-\vb*{q},l}+h.c.\notag\\
\mathcal{H}_{\text{int}_2}&=\sum_{\vb*{k},\vb*{q},\vb*{p}}\sum_{\mu}\sum_{l}\frac{\tilde{W}^{\mu}}{N}\notag\\
&\times\phi^\dagger_{\text{e},\vb*{k},\mu,l}\phi_{\text{e},\vb*{q},\mu,l}\phi^\dagger_{\text{m},\vb*{p},l}\phi_{\text{m},\vb*{k}-\vb*{q}+\vb*{p},l},
\end{align}
\end{subequations}
where $l$ is the sublattice index ($A$ or $B$), $\tilde{V}=-J_{ex}\sqrt{2S}$, $\tilde{W}^{\uparrow}=J_{ex}$, and $\tilde{W}^{\downarrow}=-J_{ex}$.
In this model, the electron Hamiltonian conserves spin, and we can rewrite the interaction Hamiltonian in the band basis as
\begin{subequations}
\begin{align}
\mathcal{H}_{\text{int}_1}=& \sum_{\vb*{k},\vb*{q}}\sum_{\alpha,\beta,\gamma}\frac{V^{\alpha\beta\gamma}}{\sqrt{N}}\psi^\dagger_{\text{e},\vb*{k},\downarrow,\alpha}\psi_{\text{e},\vb*{q},\uparrow,\beta}\psi_{\text{m},\vb*{k}-\vb*{q},\gamma}+h.c.\notag\\
\mathcal{H}_{\text{int}_2}&=\sum_{\vb*{k},\vb*{q},\vb*{p}}\sum_{\mu}\sum_{\alpha,\beta,\gamma,\delta}\frac{W^{\mu,\alpha\beta\gamma\delta}}{N}\notag\\
&\times\psi^\dagger_{\text{e},\vb*{k},\mu,\alpha}\psi_{\text{e},\vb*{q},\mu,\beta}\psi^\dagger_{\text{m},\vb*{p},\gamma}\psi_{\text{m},\vb*{k}-\vb*{q}+\vb*{p},\delta},
\end{align}
\end{subequations}
where $\alpha$, $\beta$, $\gamma$, and $\delta$ are the band indices and this is the form in TMDs discussed in the main text.

In general, due to spin-orbit coupling, the electron Hamiltonian does not conserve spin. Consequently, spin is no longer a good quantum number, and the interaction Hamiltonian in the band basis is given by
\begin{subequations}
\begin{align}
\mathcal{H}_{\text{int}_1}=& \sum_{\vb*{k},\vb*{q}}\sum_{\alpha,\beta,\gamma}\frac{V_{\vb*{k},\vb*{q}}^{\alpha\beta\gamma}}{\sqrt{N}}\psi^\dagger_{\text{e},\vb*{k},\alpha}\psi_{\text{e},\vb*{q},\beta}\psi_{\text{m},\vb*{k}-\vb*{q},\gamma}+h.c.\notag\\
\mathcal{H}_{\text{int}_2}&=\sum_{\vb*{k},\vb*{q},\vb*{p}}\sum_{\alpha,\beta,\gamma,\delta}\frac{W_{\vb*{k},\vb*{q},\vb*{p}}^{\alpha\beta\gamma\delta}}{N}\notag\\
&\times\psi^\dagger_{\text{e},\vb*{k},\alpha}\psi_{\text{e},\vb*{q},\beta}\psi^\dagger_{\text{m},\vb*{p},\gamma}\psi_{\text{m},\vb*{k}-\vb*{q}+\vb*{p},\delta}.\label{eq:general_int}
\end{align}
\end{subequations}
This corresponds to the form for graphene with Rashba SOC discussed in Appendix~\ref{sec:appendix_graphene}.

\section{Derivation of effective Hamiltonian} \label{sec:appendix_effective_hamiltonian}
Here, we derive the effective Hamiltonian. The total Hamiltonian is given by
\begin{equation}
    \mathcal{H} = \mathcal{H}_{\text{m}} + \mathcal{H}_{\text{e}}^\prime + \mathcal{H}_{\text{int}_1} + \mathcal{H}_{\text{int}_2},
\end{equation}
and we consider the general form of the interaction Hamiltonian given in Eq.~\eqref{eq:general_int}.

By replacing the electron operators in $\mathcal{H}_{\textrm{int}_2}$ with their expectation values for the non-interacting electron system, we obtain the effective interaction between magnons derived from $\mathcal{H}_{\textrm{int}_2}$ as
\begin{equation}
    \mathcal{H}_{\textrm{int}_2} = \frac{1}{N}\sum_{\vb*{q}}\sum_{\gamma}W^{\gamma\gamma\alpha\beta}f_F(\varepsilon_{\vb*{q},\gamma})\psi^\dagger_{\text{m},\vb*{k},\alpha}\psi_{\text{m},\vb*{k},\beta}.
\end{equation}
To obtain the effective Hamiltonian from $H_{\textrm{int}_1}$, we use the Schrieffer-Wolff transformation.
We define the effective Hamiltonian via the unitary transformation
\begin{equation}
    \mathcal{H}_{\text{eff}} = e^{S}(\mathcal{H}_{\textrm{m}} + \mathcal{H}_{\textrm{e}}^\prime + \mathcal{H}_{\textrm{int}1})e^{-S}.
\end{equation}
By using the Baker-Campbell-Hausdorff formula, the effective Hamiltonian is expanded as
\begin{equation}
    \mathcal{H}_{\text{eff}} = \mathcal{H}_{\textrm{m}} + \mathcal{H}_{\textrm{e}}^\prime + \mathcal{H}_{\textrm{int}1} + [S,\mathcal{H}_{\textrm{m}}+\mathcal{H}_{\textrm{e}}^\prime] + [S,\mathcal{H}_{\textrm{int}1}] +\cdots.
\end{equation}
We choose $S$ such that it satisfies $\mathcal{H}_{\textrm{int}1} + [S,\mathcal{H}_{\textrm{m}}+\mathcal{H}_{\textrm{e}}^\prime]=0$. Under this condition, the effective Hamiltonian is given by
\begin{equation}
    \mathcal{H}_{\text{eff}} = \mathcal{H}_e^\prime + \mathcal{H}_m + \frac{1}{2}[S,\mathcal{H}_{\text{int}}].
\end{equation}
The operator $S$ which satisfies $\mathcal{H}_{\textrm{int}1} + [S,\mathcal{H}_{\textrm{m}}+\mathcal{H}_{\textrm{e}}^\prime]=0$ is given by
\begin{align}
    S = &\frac{-1}{\sqrt{N}}\sum_{\vb*{k},\vb*{q}}\sum_{\alpha\beta\gamma}\frac{V_{\vb*{k},\vb*{q}}^{\alpha\beta\gamma}}{\omega_{\vb*{k}-\vb*{q},\gamma}-\varepsilon_{\vb*{k},\alpha}+\varepsilon_{\vb*{q},\beta}}\notag\\
    &\times\psi^\dagger_{\text{e},\vb*{k},\alpha}\psi_{\text{e},\vb*{q},\beta}\psi_{\text{m},\vb*{k}-\vb*{q},\gamma}\notag\\
    &+\frac{1}{\sqrt{N}}\sum_{\vb*{k},\vb*{q}}\sum_{\alpha\beta\gamma}\frac{(V_{\vb*{k},\vb*{q}}^{\alpha\beta\gamma})^*}{\omega_{\vb*{k}-\vb*{q},\gamma}-\varepsilon_{\vb*{k},\alpha}+\varepsilon_{\vb*{q},\beta}}\notag\\
    &\times\psi^\dagger_{\text{e},\vb*{q},\beta}\psi_{\text{e},\vb*{k},\alpha}\psi^\dagger_{\text{m},\vb*{k}-\vb*{q},\gamma}.
\end{align}
Therefore, the commutator $[S,\mathcal{H}_{\text{int}_1}]$ is given by
\begin{align}
    [S,\mathcal{H}_{\text{int}_1}] =& \frac{-1}{N}\sum_{\vb*{k},\vb*{q},\vb*{p}}\sum_{\alpha,\beta,\gamma,\delta,\eta}\bigg[\notag\\
    &\frac{V_{\vb*{k},\vb*{q}}^{\alpha\beta\gamma}(V_{\vb*{p},\vb*{q}}^{\eta\beta\delta})^*}{\omega_{\vb*{k}-\vb*{q},\gamma}-\varepsilon_{\vb*{k},\alpha}+\varepsilon_{\vb*{q},\beta}}\psi^\dagger_{\text{e},\vb*{k},\alpha}\psi_{\text{e},\vb*{p},\eta}\notag\\
    &\times(\psi_{\text{m},\vb*{k}-\vb*{q},\gamma}\psi^\dagger_{\text{m},\vb*{p}-\vb*{q},\delta}+\psi^\dagger_{\text{m},\vb*{k}-\vb*{q},\gamma}\psi_{\text{m},\vb*{p}-\vb*{q},\delta})\notag\\
    &-\frac{V_{\vb*{k},\vb*{q}}^{\alpha\beta\gamma}(V_{\vb*{k},\vb*{p}}^{\alpha\eta\delta})^*}{\omega_{\vb*{k}-\vb*{q},\gamma}-\varepsilon_{\vb*{k},\alpha}+\varepsilon_{\vb*{q},\beta}}\psi^\dagger_{\text{e},\vb*{k},\eta}\psi_{\text{e},\vb*{p},\beta}\notag\\
    &\times(\psi_{\text{m},\vb*{k}-\vb*{q},\gamma}\psi^\dagger_{\text{m},\vb*{k}-\vb*{p},\delta}+\psi^\dagger_{\text{m},\vb*{k}-\vb*{q},\gamma}\psi_{\text{m},\vb*{k}-\vb*{p},\delta})\notag\\
    &+\bigg\{\frac{V_{\vb*{k},\vb*{q}}^{\alpha\beta\gamma}V_{\vb*{q},\vb*{p}}^{\beta\eta\delta}}{\omega_{\vb*{k}-\vb*{q},\gamma}-\varepsilon_{\vb*{k},\alpha}+\varepsilon_{\vb*{q},\beta}}\notag\\
    &\times\psi^\dagger_{\text{e},\vb*{k},\alpha}\psi_{\text{e},\vb*{p},\eta}\psi_{\text{m},\vb*{k}-\vb*{q},\gamma}\psi_{\text{m},\vb*{q}-\vb*{p},\delta}\notag\\
    &-\frac{V_{\vb*{k},\vb*{q}}^{\alpha\beta\gamma}(V_{\vb*{p},\vb*{k}}^{\eta\alpha\delta})^*}{\omega_{\vb*{k}-\vb*{q},\gamma}-\varepsilon_{\vb*{k},\alpha}+\varepsilon_{\vb*{q},\beta}}\notag\\
    &\times\psi^\dagger_{\text{e},\vb*{k},\eta}\psi_{\text{e},\vb*{p},\beta}\psi_{\text{m},\vb*{k}-\vb*{q},\gamma}\psi_{\text{m},\vb*{p}-\vb*{k},\delta}\bigg\}+h.c.\bigg]\notag\\
    +&\psi^\dagger_e\psi^\dagger_e\psi_e\psi_e \text{terms}.
\end{align} 
Since we consider the effective Hamiltonian of magnons, we focus on the magnon part and we ignore the last term. By replacing the electron operators $\psi^\dagger_e\psi_e$ with their expectation values with respect to the non-interacting ground state, we obtain
\begin{align}
    [S,\mathcal{H}_{\text{int}_1}] \approx& \frac{-1}{N}\sum_{\vb*{k},\vb*{q}}\sum_{\alpha,\beta,\gamma,\delta}\bigg[\frac{V_{\vb*{k},\vb*{q}}^{\alpha\beta\gamma}(V_{\vb*{k},\vb*{q}}^{\alpha\beta\delta})^*(f_F(\varepsilon_{\vb*{k},\alpha})-f_F(\varepsilon_{\vb*{q},\beta}))}{\omega_{\vb*{k}-\vb*{q},\gamma}-\varepsilon_{\vb*{k},\alpha}+\varepsilon_{\vb*{q},\beta}}\notag\\
    &\times(\psi_{\text{m},\vb*{k}-\vb*{q},\gamma}\psi^\dagger_{\text{m},\vb*{k}-\vb*{q},\delta}+\psi^\dagger_{\text{m},\vb*{k}-\vb*{q},\gamma}\psi_{\text{m},\vb*{k}-\vb*{q},\delta})\notag\\
    &+\bigg\{\frac{V_{\vb*{k},\vb*{q}}^{\alpha\beta\gamma}V_{\vb*{q},\vb*{k}}^{\beta\alpha\delta}(f_F(\varepsilon_{\vb*{k},\alpha})-f_F(\varepsilon_{\vb*{q},\beta}))}{\omega_{\vb*{k}-\vb*{q},\gamma}-\varepsilon_{\vb*{k},\alpha}+\varepsilon_{\vb*{q},\beta}}\notag\\
    &\times(\psi_{\text{m},\vb*{k}-\vb*{q},\gamma}\psi_{\text{m},\vb*{q}-\vb*{k},\delta})\bigg\}+h.c.\bigg].
\end{align}
Thus, the effective magnon Hamiltonian is given by 
\begin{equation}
    \mathcal{H}_{eff} = \sum_{\vb*{k}}\psi^\dagger_{\text{m},\vb*{k}}H_{\text{eff},\vb*{k}}\psi_{\text{m},\vb*{k}}+\sum_{\alpha,\beta}(H^{\alpha\beta}_{BdG,\vb*{k}}\psi_{\text{m},\vb*{k},\alpha}\psi_{\text{m},-\vb*{k},\beta}+h.c.),
\end{equation}
where 
\begin{align}
    H_{\text{eff},\vb*{k}}^{\alpha\beta} =& H_{\text{m},\vb*{k}}^{\alpha,\beta}+\frac{1}{N}\sum_{\vb*{q}}\sum_{\gamma}W^{\gamma\gamma\alpha\beta}f_F(\varepsilon_{\vb*{q},\gamma})\notag\\
    &-\frac{1}{2N}\sum_{\vb*{q}}\sum_{\gamma,\delta} (V^{\gamma\delta\alpha}_{\vb*{q},\vb*{q}-\vb*{k}})^*V^{\gamma\delta\beta}_{\vb*{q},\vb*{q}-\vb*{k}}\notag\\
    &\times\frac{1}{2}[\frac{(f_F(\varepsilon_{\vb*{q},\gamma})-f_F(\varepsilon_{\vb*{q}-\vb*{k},\delta}))}{\omega_{\vb*{k},\beta}-\varepsilon_{\vb*{q},\gamma}+\varepsilon_{\vb*{q}-\vb*{k},\delta}}\notag\\
    &+\frac{(f_F(\varepsilon_{\vb*{q},\gamma})-f_F(\varepsilon_{\vb*{q}-\vb*{k},\delta}))}{\omega_{\vb*{k},\alpha}-\varepsilon_{\vb*{q},\gamma}+\varepsilon_{\vb*{q}-\vb*{k},\delta}}],\label{eq:Heff_general}
\end{align}
and
\begin{align}
    H^{\alpha\beta}_{BdG,\vb*{k}}=& -\frac{1}{2N}\sum_{\vb*{q}}\sum_{\gamma,\delta} V^{\gamma\delta\alpha}_{\vb*{q},\vb*{q}-\vb*{k}}V^{\delta\gamma\beta}_{\vb*{q}-\vb*{k},\vb*{q}}\notag\\
    &\times\frac{(f_F(\varepsilon_{\vb*{q},\gamma})-f_F(\varepsilon_{\vb*{q}-\vb*{k},\delta}))}{\omega_{\vb*{k},\alpha}+\omega_{\vb*{k},\beta}-\varepsilon_{\vb*{q},\gamma}+\varepsilon_{\vb*{q}-\vb*{k},\delta}}.\label{eq:BdG_general}
\end{align}
In particular, when the electron system conserves spin, as in the case of TMDs, $\gamma$ and $\delta$ in $V^{\gamma\delta\alpha}_{\vb*{q},\vb*{q}-\vb*{k}}$ represent band indices of spin-down and spin-up electrons, respectively. Therefore, when $V^{\gamma\delta\alpha}_{\vb*{q},\vb*{q}-\vb*{k}}$ is non-zero, $V^{\delta\gamma\beta}_{\vb*{q}-\vb*{k},\vb*{q}}$ is zero, and vice versa.
Thus, the BdG term vanishes in this case and the effective magnon Hamiltonian is given by
\begin{align}
    H_{\text{eff},\vb*{k}}^{\alpha\beta} =& \omega_{\vb*{k}}^{\alpha\beta}+\frac{1}{N}\sum_{\vb*{q}}\sum_{\gamma}W^{\mu,\gamma\gamma\alpha\beta}f_F(\varepsilon_{\vb*{q},\mu,\gamma})\notag\\
    &-\frac{1}{2N}\sum_{\vb*{q}}\sum_{\gamma,\delta} V_{\vb*{q},\vb*{q}-\vb*{k}}^{\gamma\delta\beta}(V_{\vb*{q},\vb*{q}-\vb*{k}}^{\gamma\delta\alpha})^*\notag\\
    &\times\frac{1}{2}\bigg[\frac{(f_F(\varepsilon_{\vb*{q},\gamma})-f_F(\varepsilon_{\vb*{q}-\vb*{k},\delta}))}{\omega_{\vb*{k},\beta}-\varepsilon_{\vb*{q},\downarrow,\gamma}+\varepsilon_{\vb*{q}-\vb*{k},\uparrow,\delta}}+(\alpha\leftrightarrow\beta)\bigg].
\end{align}
This corresponds to the effective magnon Hamiltonian for TMDs discussed in the main text.

\section{Green's function and spectral function}\label{sec:appendix_spectral}
Here, we consider the renormalized magnon Green's function $\mathcal{G}_m$ and spectral function in the presence of electron-magnon interactions.
The unperturbed Matsubara Green's function of a magnon in the imaginary-time formalism $\mathcal{G}^0_{\text{m},\vb*{k},\alpha\beta}(\tau)$ is given as
\begin{align}
    \mathcal{G}^0_{\text{m},\vb*{k},\alpha\beta}(\tau)=&-\langle T_\tau \psi_{\text{m},\vb*{k},\alpha}(\tau)\psi^\dagger_{\text{m},\vb*{k},\beta}\rangle,
\end{align}
where $T_\tau$ represents the imaginary time-ordering operators, and $\langle\cdots\rangle$ is the thermal average with respect to the unperturbed Hamiltonian $\mathcal{H}_m$.
By performing a Fourier transform to the Matsubara frequency $\omega_n=2n\pi k_BT$, we obtain
\begin{align}
    \mathcal{G}^0_{\text{m},\vb*{k},\alpha\beta}(i\omega_n) = \frac{\delta_{\alpha,\beta}}{i\omega_n-\omega_{\vb*{k},\alpha}}.
\end{align}
To incorporate the effect of electron-magnon interactions, we consider the perturbation expansion of the Green's function. 
The Green's function of magnon $\mathcal{G}_m$ in the imaginary time formalism is given by
\begin{align}
    &\mathcal{G}_{\text{m},\vb*{k},\alpha\beta}(\tau)=\notag\\
    &\mathcal{G}_{\text{m},\vb*{k},\alpha\beta}^0(\tau)+\int^\beta_0d\tau_1\langle T_\tau \mathcal{H}_{\text{int}_2}(\tau_1)\psi_{\text{m},\vb*{k},\alpha}(\tau)\psi^\dagger_{\text{m},\vb*{k},\beta}\rangle\notag\\
    &-\frac{1}{2}\int^\beta_0d\tau_1\int^\beta_0d\tau_2\langle T_\tau \mathcal{H}_{\text{int}_1}(\tau_1)\mathcal{H}_{\text{int}_1}(\tau_2)\psi_{\text{m},\vb*{k},\alpha}(\tau)\psi^\dagger_{\text{m},\vb*{k},\beta}\rangle.\label{eq:green's_func_magnon}
\end{align}
The contributions from the second and third terms on the right-hand side correspond to the self-energies  $\Sigma_{\text{cor}}$ and $\Sigma_{\text{sca}}$, respectively. The correction term $\Sigma_{\text{cor}}$ is given by
\begin{align}
    \Sigma^{\alpha\beta}_{\text{cor},\vb*{k},i\omega_n} = \frac{k_BT}{N}\sum_{\vb*{q}}\sum_{\mu}\sum_{\gamma\delta}\sum_m W_{\mu,\vb*{q},\vb*{q},\vb*{k}}^{\gamma\delta\alpha\beta}\mathcal{G}_{\text{e},\vb*{q},\gamma\delta}(i\varepsilon_m),
\end{align}
where $\mathcal{G}_e(i\varepsilon_m)$ is the Matsubara Green's function of an electron with a Matsubara frequency $\varepsilon_m=(2m+1)\pi k_BT$. We calculate $\Sigma_{\text{cor}}$ by using the unperturbed Green's function of electron 
\begin{equation}
    \mathcal{G}_{\text{e},\vb*{k},\alpha\beta}(i\varepsilon_n) = \frac{\delta_{\alpha,\beta}}{i\varepsilon_n-\varepsilon_{\vb*{k},\alpha}},
\end{equation}
and we obtain
\begin{equation}
    \Sigma_{\text{cor},\vb*{k},\omega}^{\alpha,\beta}= \frac{1}{N}\sum_{\vb*{q}}\sum_{\gamma} W_{\vb*{q},\vb*{q},\vb*{k}}^{\gamma\gamma\alpha\beta}f_F(\varepsilon_{\vb*{q},\gamma})\delta_{\alpha,\beta},\label{General_Sigma_cor}.
\end{equation}
Thus, the self-energy $\Sigma_{\text{cor}}$ gives the uniform energy shift of magnon bands. In particular, when the Rashba-type SOC is absent in this model, $\Sigma_{\text{cor}}$ simplifies to
\begin{equation}
    \Sigma_{\text{cor},\vb*{k},\omega}^{\alpha,\beta}= \frac{J_{ex}}{N}\sum_{\vb*{q}}\sum_{\gamma}(\Braket{n_{\vb*{q},\uparrow,\gamma}}-\Braket{n_{\vb*{q},\downarrow,\gamma}})\delta_{\alpha,\beta}.
\end{equation}

\begin{figure}[tb]
    \centering
    \includegraphics[width=\linewidth]{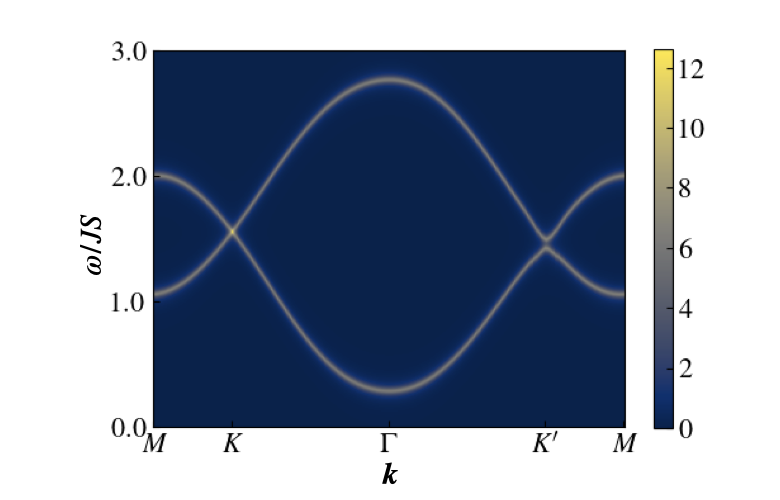}
    \caption{The magnon spectral function along the high-symmetry lines in the Brillouin zone. The parameters are the same as those in Fig.~2(b) in the main text: $t_2/t_1=0.12$, $\Delta/t_1=0.75$, $\phi=0.3\pi$, $J/t_1=1.0\times10^{-3}$, $\Delta_z/t_1=1.0\times10^{-4}$, $J_{ex}/t_1=0.07$, $S=1$, and $\varepsilon_F/t_1=0.5$.}
    \label{fig:spectral_function}
\end{figure}

The self-energy $\Sigma_{\text{sca}}$ is given by
\begin{align}
    \Sigma^{\alpha\beta}_{\text{sca},\vb*{k},i\omega_n} =& \frac{k_BT}{2N}\sum_{\vb*{q}}\sum_{\gamma,\delta,\gamma^\prime,\delta^\prime}\sum_{m}V_{\vb*{q},\vb*{q}-\vb*{k}}^{\gamma\delta\alpha}(V_{\vb*{q},\vb*{q}-\vb*{k}}^{\gamma^\prime\delta^\prime\beta})^*\notag\\
    &\times\mathcal{G}_{\text{e},\vb*{q},\gamma\gamma^\prime}(i\varepsilon_m)\mathcal{G}_{\text{e},\vb*{q}-\vb*{k},\delta\delta^\prime}(i\varepsilon_m-i\omega_n).\notag\\
\end{align}
By using the unperturbed electron Green's function, we obtain
\begin{align}
    \Sigma_{\text{sca},\vb*{k},\omega}^{\alpha,\beta}=& -\frac{1}{2N}\sum_{\vb*{q}}\sum_{\gamma,\delta} V_{\vb*{q},\vb*{q}-\vb*{k}}^{\gamma\delta\alpha}(V_{\vb*{q},\vb*{q}-\vb*{k}}^{\gamma\delta\beta})^*\notag\\
    &\times\frac{(f_F(\varepsilon_{\vb*{q},\gamma})-f_F(\varepsilon_{\vb*{q}-\vb*{k},\delta}))}{\omega+i\eta-\varepsilon_{\vb*{q},\gamma}+\varepsilon_{\vb*{q}-\vb*{k},\delta}}.\label{General_Sigma_sca}
\end{align}
This self-energy induces a nontrivial momentum dependence in the magnon bands.
The total self-energy $\Sigma_{\vb*{k},\omega}$ is given by $\Sigma_{\vb*{k},\omega} = \Sigma_{\text{cor},\vb*{k},\omega} + \Sigma_{\text{sca},\vb*{k},\omega}$ and the renormalized retarded magnon Green's function $\mathcal{G}^R_\text{m}$ is determined by the Dyson equation as
\begin{equation}
    \mathcal{G}^R_{\text{m},\vb*{k},\omega} = [\omega + i\eta - H_{\text{m},\vb*{k}} - \Sigma_{\vb*{k},\omega}]^{-1}.
\end{equation}
The magnon spectral function $A_{\vb*{k},\omega}$ is defined as
\begin{equation}
    A_{\vb*{k},\omega} =-\frac{1}{\pi}\text{Im}\{\text{Tr}[\mathcal{G}^R_{\text{m},\vb*{k},\omega}]\}.
\end{equation}
The magnon spectral function reveals the renormalized band structure in the presence of electron-magnon interactions. Figure~\ref{fig:spectral_function} shows the magnon spectral function along high-symmetry lines in the Brillouin zone. The parameters used are the same as those in Fig.~\ref{fig:TMD_magnon}(a) of the main text. The spectral function shows the nontrivial band deformation of magnon bands due to the electron-magnon interaction. In particular, the gap opens at the $K$ and $K^\prime$ points. The spectral function is in good agreement with the magnon band structure obtained from the effective magnon Hamiltonian shown in Fig.~\ref{fig:TMD_magnon}(a) in the main text. 

\section{Effective magnon mass of $K^\prime$ valley}
\label{sec:appendix_valley}

\begin{figure}[tb]
    \centering
    \includegraphics[width=\linewidth]{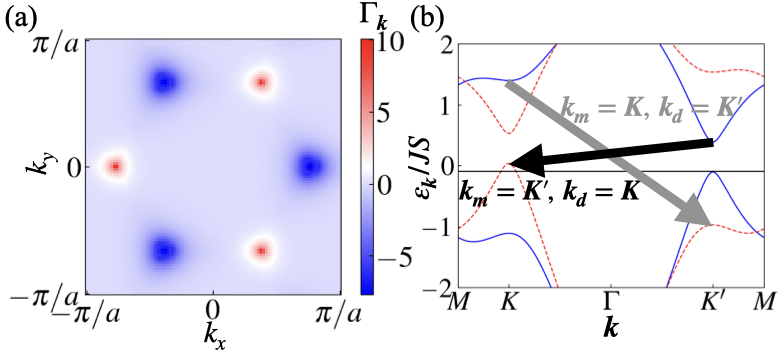}
    \caption{Berry curvature and schematics of transition processes. (a) Berry curvature of the lowest magnon band when the Fermi energy intersects the spin-down bands around the $K$ valley ($\varepsilon_F/t_1=-0.1$). (b) Schematics of transition processes ($\varepsilon_F/t_1=-0.1$).
    We use the following parameters: $t_2/t_1=0.12$, $\Delta/t_1=0.75$, $\phi=0.3\pi$, $J/t_1=1.0\times10^{-3}$, $\Delta_z/t_1=1.0\times10^{-4}$, $J_{ex}/t_1=0.07$, and $S=1$.}
    \label{fig:valley_mass}
\end{figure}

In this section, we consider the Berry curvature when the Fermi energy intersects the spin-down bands around the $K$ valley ($\varepsilon_F/t_1=-0.1$). Figure~\ref{fig:valley_mass}(a) shows the Berry curvature of the lowest magnon band in this case.
The Berry curvature around the $K$ point is positive, similar to the result shown in Fig.~\ref{fig:TMD_magnon}(b) in the main text. In contrast, the Berry curvature around the $K^\prime$ point is negative, opposite to that shown in Fig.~\ref{fig:TMD_magnon}(b). This can be understood from the schematics of transition processes shown in Fig.~\ref{fig:valley_mass}(b). When the Fermi energy intersects the spin-down bands around the $K$ valley, the dominant contribution to the effective magnon Hamiltonian at the $K$ point is the same as that shown in Fig.~\ref{fig:TMD_magnon}(c). Therefore, the Berry curvature around the $K$ point is positive. Conversely, the dominant contribution to the effective magnon Hamiltonian at the $K^\prime$ point is different from that shown in Fig.~\ref{fig:TMD_magnon}(c), resulting in a negative Berry curvature around the $K^\prime$ point. 
Consequently, the Chern number of the lowest magnon band is zero in this case because the contributions from the $K$ and $K^\prime$ points cancel out.

\section{Graphene with Rashba SOC} \label{sec:appendix_graphene}
Now, we consider graphene with Rashba SOC coupled to a ferromagnetic layer via the proximity effect. The spin and interaction Hamiltonians are the same as those in Eqs.~\eqref{eq:spin_hamiltonian} and \eqref{eq:interaction}. 

The electron Hamiltonian is given by
\begin{subequations}
\begin{align}
    \mathcal{H}_{\text{e}}&=\sum_{\vb*{k}}\phi^\dagger_{\text{e},\vb*{k}}H_{\text{e},\vb*{k}}\phi_{\text{e},\vb*{k}}\\
    \phi_{\text{e},\vb*{k}}&=(c_{\vb*{k},\uparrow,A},c_{\vb*{k},\uparrow,B},c_{\vb*{k},\downarrow,A},c_{\vb*{k},\downarrow,B})^T,
\end{align}
\end{subequations}
and $H_{\text{e},\vb*{k}}$ is given by
\begin{equation}
    H_{\text{e},\vb*{k}} = \begin{pmatrix}
        -t_2f_{\vb*{k}}+\Delta & -t_1\gamma_{\vb*{k}} & 0 & \lambda_{+,\vb*{k}}\\
        -t_1\gamma_{\vb*{k}}^* & -t_2f_{\vb*{k}}-\Delta & \lambda_{-,\vb*{k}} & 0\\
        0 & \lambda_{-,\vb*{k}}^* & -t_2f_{\vb*{k}}+\Delta & -t_1\gamma_{\vb*{k}}\\
        \lambda_{+,\vb*{k}}^* & 0 & -t_1\gamma_{\vb*{k}}^* & -t_2f_{\vb*{k}}-\Delta
        \end{pmatrix},
\end{equation}
where $f_{\vb*{k}}=f(\phi=0,\vb*{k})$, $\lambda_{\pm,\vb*{k}}=\pm i\lambda_R\sum_{l}(i\delta_{l,x}+\delta_{l,y})e^{\pm i\vb*{k}\cdot\vb*{\delta}_l}$, and $\lambda_R$ is the Rashba SOC strength.
This Rashba SOC arises from the broken mirror symmetry along the z-axis due to the presence of the ferromagnetic layer.

\begin{figure}[htb]
    \centering
    \includegraphics[width=\linewidth]{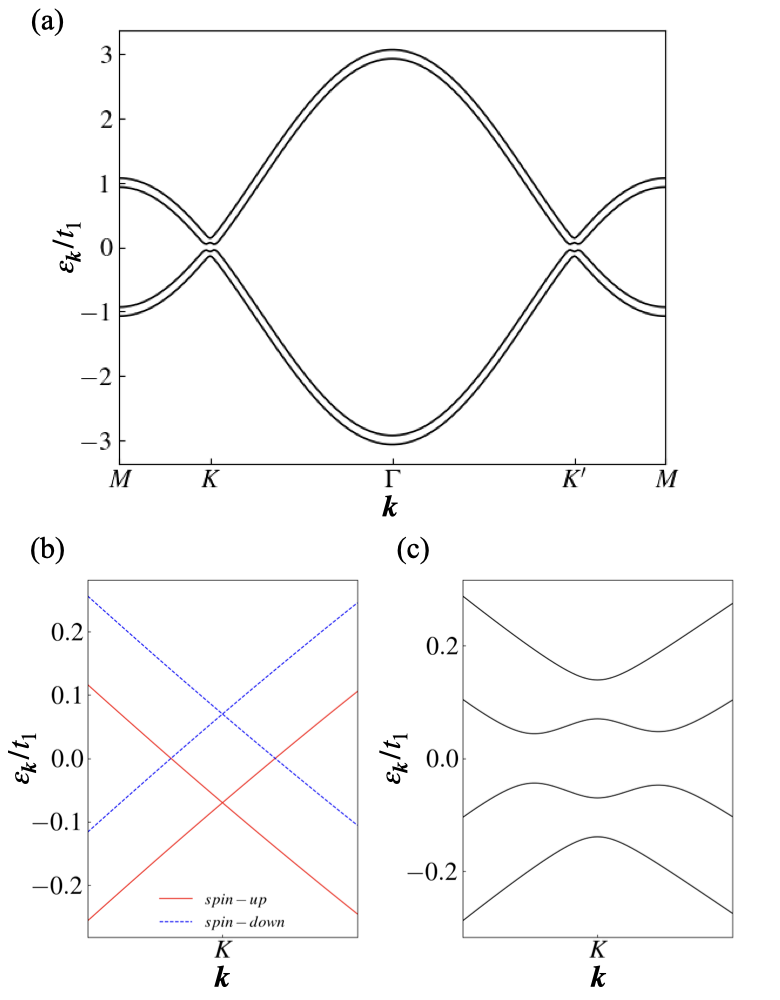}
    \caption{The band structure of the electron Hamiltonian. (a) The band structure with finite SOC ($\lambda_R/t_1=0.04$). (b) The band structure around the $K$ point without SOC. (c) The band structure around the $K$ point with finite SOC ($\lambda_R/t_1=0.04$). We use the following parameters: $t_2/t_1=0.0$, $\Delta/t_1=0.0$, and $J_{ex}/t_1=0.07$.} 
    \label{fig:band}
\end{figure}

Due to the proximity effect of the ferromagnetic layer, the electron system acquires an effective exchange field. Therefore, the electron Hamiltonian includes the effective exchange field induced by the interaction Eq.\eqref{eq:interaction}. The effective exchange field term is explicitly given by $-J_{ex}S\sigma_z\otimes\tau_0$. The band structure of the electron Hamiltonian in the presence of this field is shown in Fig.~\ref{fig:band}. When Rashba SOC is finite, a gap opens at the $K$ and $K^\prime$ points, as shown in Fig.~\ref{fig:band}(a). The band structure around the $K$ point is shown in Fig.~\ref{fig:band}(b) and (c). In the absence of Rashba SOC, the bands remain gapless as shown in Fig.~\ref{fig:band}(b). However, when Rashba SOC is finite, spin-up and spin-down bands hybridize, opening a gap around the $K$ point as shown in Fig.~\ref{fig:band}(c). 

\begin{figure}[htb]
    \centering
    \includegraphics[width=0.9\linewidth]{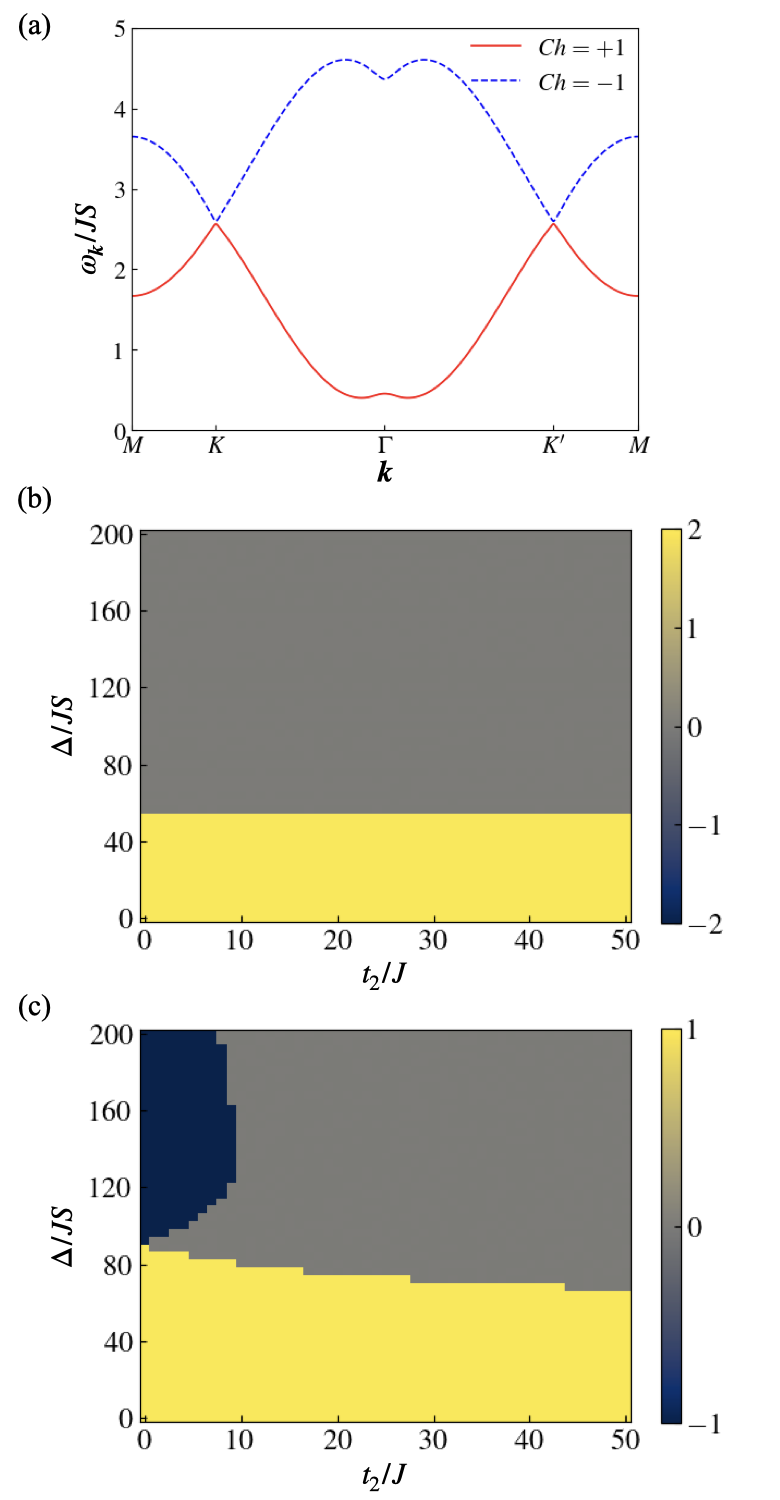}
    \caption{The band structure and phase diagram of the effective Hamiltonian. (a) The band structure and Chern number of the effective Hamiltonian. (b) The phase diagram of the electron Hamiltonian. We plot the Chern number of occupied bands. (c) The phase diagram of the effective magnon Hamiltonian. We plot the Chern number of the lowest band.    We use the following parameters: $t_2/t_1=0.0$, $\lambda_R/t_1=0.04$, $\varepsilon_F=3t_2$, $J/t_1=1.0\times10^{-3}$, $\Delta_z/t_1=1.0\times10^{-4}$, $J_{ex}/t_1=0.07$, and $S=1$.} 
    \label{fig:graphene_magnon}
\end{figure}

Since this model does not conserve electron spin due to the Rashba SOC, we use the general form of interaction Hamiltonian given in Eq.~\eqref{eq:general_int} and the effective magnon Hamiltonian is given by Eq.~\eqref{eq:Heff_general} and Eq.~\eqref{eq:BdG_general} in Appendix~\ref{sec:appendix_effective_hamiltonian}.
The band structure of the effective magnon Hamiltonian is shown in Fig.~\ref{fig:graphene_magnon}. 
The magnon bands acquire nontrivial Chern numbers due to electron-magnon interactions and the associated gap opening at the $K$ and $K^\prime$ points. A calculation of the Chern numbers for the effective magnon bands confirms that they are topological magnons.

To check the robustness of the magnon topology, we consider the staggered potential $\Delta$, which breaks the inversion symmetry of the electron system. As shown in Fig.~\ref{fig:graphene_magnon}(b), the electron system becomes topologically trivial when the staggered potential is sufficiently large. (Note that we set the Fermi energy to $\varepsilon_F=3t_2$ to ensure it lies within the band gap).
Remarkably, the effective magnon Hamiltonian remains topological even in the regime where the electron system is trivial, as shown in Fig.~\ref{fig:graphene_magnon} (c). In particular, when the next-nearest-neighbor hopping is absent ($t_2=0$), the effective magnon Hamiltonian is topological, regardless of the size of the staggered potential $\Delta$.

This robustness at $t_2=0$ is protected by a specific symmetry of the total Hamiltonian. We consider a combined operation of the spatial inversion for magnons, $P_m$ and the anti-unitary operation acting on electrons, $UK$. Although the staggered potential breaks the inversion symmetry of the electron system, the total Hamiltonian remains invariant under the combined transformation when $t_2=0$ (see Appendix~\ref{sec:appendix_symmetry} for the explicit derivation). Consequently, the effective magnon Hamiltonian preserves the inversion symmetry when $t_2=0$. When $t_2\neq0$, this symmetry is broken. However, the symmetry breaking effect transmitted to the magnon system is weak. Thus, the effective magnon Hamiltonian can remain topological even if the staggered potential $\Delta$ is large and the electron Hamiltonian is trivial.

Finally, we estimate the magnitude of the topological magnon gap. Using the parameters in Fig.~\ref{fig:graphene_magnon}, the gap size is of the order of $10^{-2}JS$ and the magnitude of the topological gap is proportional to $J_{ex}^2S/\varepsilon_e$ as discussed in the main text. Here, $\varepsilon_e$ is the energy scale of the electron system, which is of the order of the transfer integral $t_1$ in graphene. Therefore, in the electron system with large transfer integral $t_1$, such as graphene, the topological magnon gap is of the order of $10^{-3}JS$. However, in the electron system with small transfer integral $t_1$, we can expect a larger topological gap. Thus, electron systems with small transfer integral $t_1$, such as twisted bilayer graphene at the magic angle, are promising candidates for observing large topological magnon gaps induced by electron-magnon interactions.

\section{Symmetry of total Hamiltonian}\label{sec:appendix_symmetry}
Here, we discuss the symmetry of the total Hamiltonian. We focus on the magnon with momentum $-\vb*{k}$. The total Hamiltonian associated with the magnon with momentum $-\vb*{k}$ $H_{\text{tot},-\vb*{k}}$ is
\begin{align}
    H_{\text{tot},-\vb*{k}}=&\frac{1}{2}\phi_{\text{m},-\vb*{k}}^\dagger H_{\text{m},-\vb*{k}}\phi_{\text{m},-\vb*{k}}+\sum_{\vb*{q}}\phi_{\text{e},\vb*{q}}^\dagger H_{\text{e},\vb*{q}}\phi_{\text{e},\vb*{q}}\notag\\
    &-\frac{J_{ex}\sqrt{2S}}{\sqrt{N}}\sum_{\vb*{q}}\sum_{l}c^\dagger_{\vb*{q}-\vb*{k},\downarrow,l}c_{\vb*{q},\uparrow,l}\phi_{\text{m},-\vb*{k},l}+h.c.\notag\\
    &+\frac{J_{ex}}{N}\sum_{\vb*{q}}\sum_{l}(c^\dagger_{\vb*{q},\uparrow,l}c_{\vb*{p},\uparrow,l}-c^\dagger_{\vb*{q},\downarrow,l}c_{\vb*{p},\downarrow,l})\notag\\
    &\times\phi^\dagger_{\text{m},-\vb*{k},l}\phi_{\text{m},-\vb*{k}+\vb*{q}-\vb*{p},l},
\end{align}
where $l$ denotes the sublattice index ($A$ or $B$) and operators $\phi_{\text{m},\vb*{k},A}$ and $\phi_{\text{m},\vb*{k},B}$ correspond to the magnon annihilation operators $a_{\vb*{k}}$ and $b_{\vb*{k}}$, respectively.
We consider the inversion operation of magnon $P_m$ and the anti-unitary operation of electron $UK$. The inversion operation of magnon $P_m$ is defined as
\begin{subequations}
\begin{align}
    P_ma_{\vb*{k}}P_m^{-1} &= b_{-\vb*{k}},\\
    P_mb_{\vb*{k}}P_m^{-1} &= a_{-\vb*{k}},
\end{align}
\end{subequations}
and the magnon Hamiltonian $H_{\text{m},\vb*{k}}$ is invariant under this operation since we consider the inversion-symmetric spin Hamiltonian in Eq.~\eqref{eq:spin_hamiltonian}.
The anti-unitary operation of electron $UK$ is defined as
\begin{subequations}
\begin{align}
    UKc_{\vb*{k},A,\uparrow}UK^{-1} &= ic^\dagger_{\vb*{k},B,\downarrow},\\
    UKc_{\vb*{k},A,\downarrow}UK^{-1} &= -ic^\dagger_{\vb*{k},B,\uparrow},\\
    UKc_{\vb*{k},B,\uparrow}UK^{-1} &= ic^\dagger_{\vb*{k},A,\downarrow},\\
    UKc_{\vb*{k},B,\downarrow}UK^{-1} &= -ic^\dagger_{\vb*{k},A,\uparrow}.
\end{align}
\end{subequations}
When the next-nearest-neighbor hopping $t_2$ is absent, the electron Hamiltonian $H_{\text{e},\vb*{k}}$ is invariant under this operation.
Therefore, due to operations $P_m$ and $UK$, the interaction between electrons and magnons $\mathcal{H}_{\text{int}_1}$ for $A$ sublattice transforms as
\begin{align}
    &-\frac{J_{ex}\sqrt{2S}}{\sqrt{N}}\sum_{\vb*{k},\vb*{q}}c^\dagger_{\vb*{k},A,\downarrow}c_{\vb*{q},A,\uparrow}a_{\vb*{k}-\vb*{q}}+h.c.\notag\\
    &\rightarrow -\frac{J_{ex}\sqrt{2S}}{\sqrt{N}}\sum_{\vb*{k},\vb*{q}}-c_{\vb*{k},B,\uparrow}c^\dagger_{\vb*{q},B,\downarrow}b_{-\vb*{k}+\vb*{q}}+h.c.\notag\\
    &= -\frac{J_{ex}\sqrt{2S}}{\sqrt{N}}\sum_{\vb*{k},\vb*{q}}c^\dagger_{\vb*{q},B,\downarrow}c_{\vb*{k},B,\uparrow}b_{-\vb*{k}+\vb*{q}}+h.c.\notag\\
    &= -\frac{J_{ex}\sqrt{2S}}{\sqrt{N}}\sum_{\vb*{k},\vb*{q}}c^\dagger_{\vb*{k},B,\downarrow}c_{\vb*{q},B,\uparrow}b_{\vb*{k}-\vb*{q}}+h.c.
\end{align}
In the last step, we relabeled the dummy variables $\vb*{k}$ and $\vb*{q}$.
Thus, the interaction $\mathcal{H}_{\text{int}_1}$ for the $A$ sublattice maps onto that of the $B$ sublattice. Similarly, the interaction $\mathcal{H}_{\text{int}_1}$ for the $B$ sublattice maps onto that of the $A$ sublattice. The interaction $\mathcal{H}_{\text{int}_2}$ for the $A$ sublattice also transforms to that for the $B$ sublattice as
\begin{align}
    &\frac{1}{N}\sum_{\vb*{k},\vb*{q},\vb*{p}}J_{ex}(c^\dagger_{\vb*{k},A,\uparrow}c_{\vb*{q},A,\uparrow}-c^\dagger_{\vb*{k},A,\downarrow}c_{\vb*{q},A,\downarrow})a^\dagger_{\vb*{p}}a_{\vb*{k}-\vb*{q}+\vb*{p}}\notag\\
    &\rightarrow \frac{1}{N}\sum_{\vb*{k},\vb*{q},\vb*{p}}J_{ex}(c_{\vb*{k},B,\downarrow}c^\dagger_{\vb*{q},B,\downarrow}-c_{\vb*{k},B,\uparrow}c^\dagger_{\vb*{q},B,\uparrow})b^\dagger_{-\vb*{p}}b_{-\vb*{k}+\vb*{q}-\vb*{p}}\notag\\
    &= \frac{1}{N}\sum_{\vb*{k},\vb*{q},\vb*{p}}J_{ex}(c^\dagger_{\vb*{q},B,\uparrow}c_{\vb*{k},B,\uparrow}-c^\dagger_{\vb*{q},B,\downarrow}c_{\vb*{k},B,\downarrow})b^\dagger_{-\vb*{p}}b_{-\vb*{k}+\vb*{q}-\vb*{p}}\notag\\
    &= \frac{1}{N}\sum_{\vb*{k},\vb*{q},\vb*{p}}J_{ex}(c^\dagger_{\vb*{k},B,\uparrow}c_{\vb*{q},B,\uparrow}-c^\dagger_{\vb*{k},B,\downarrow}c_{\vb*{q},B,\downarrow})b^\dagger_{\vb*{p}}b_{\vb*{k}-\vb*{q}+\vb*{p}}.
\end{align}
In the final step, we relabeled the dummy variables $\vb*{k}\leftrightarrow\vb*{q}$, and $\vb*{p}$ to $-\vb*{p}$.
The interaction $\mathcal{H}_{\text{int}_2}$ for the $B$ sublattice similarly transforms to that for the $A$ sublattice. 

Consequently, using the transformed operators $\tilde{\phi}_{\text{m},\vb*{k}}=P_m\phi_{\text{m},-\vb*{k}}$ and $\tilde{\phi}_{\text{e},\vb*{k}}=UK\phi_{\text{e},\vb*{k}}$, we find that
\begin{align}
    H_{\text{tot},-\vb*{k}}=&\frac{1}{2}\tilde{\phi}_{\text{m},\vb*{k}}^\dagger H_{\text{m},\vb*{k}}\tilde{\phi}_{\text{m},\vb*{k}}+\sum_{\vb*{q}}\tilde{\phi}_{\text{e},\vb*{q}}^\dagger H_{\text{e},\vb*{q}}\tilde{\phi}_{\text{e},\vb*{q}}\notag\\
    &+ \sum_{\vb*{q}}\sum_{\alpha,\beta,\gamma}\frac{\tilde{V}_{\vb*{q},\vb*{q}-\vb*{k}}^{\alpha\beta\gamma}}{\sqrt{N}}\tilde{\phi}^\dagger_{\text{e},\vb*{q},\alpha}\tilde{\phi}_{\text{e},\vb*{q}-\vb*{k},\beta}\tilde{\phi}_{\text{m},\vb*{k},\gamma}\notag\\
    &+\sum_{\vb*{q},\vb*{p}}\sum_{\alpha,\beta}\frac{\tilde{W}_{\vb*{q},\vb*{q},\vb*{k}}^{\alpha\alpha\beta\beta}}{N}\tilde{\phi}^\dagger_{\text{e},\vb*{q},\alpha}\tilde{\phi}_{\text{e},\vb*{p},\alpha}\tilde{\phi}^\dagger_{\text{m},\vb*{k},\beta}\tilde{\phi}_{\text{m},\vb*{k}-\vb*{q}+\vb*{p},\beta},
\end{align}
This explicitly shows that $H_{\text{tot},-\vb*{k}}$ expressed in terms of the transformed operators has the identical functional form to $H_{\text{tot},\vb*{k}}$ expressed in terms of the original operators. Therefore, by applying the derivation procedure outlined in Appendix~\ref{sec:appendix_effective_hamiltonian}, we find that the effective magnon Hamiltonian with momentum $-\vb*{k}$ has the same form as that with momentum $\vb*{k}$ when expressed in terms of the transformed operators $\tilde{\psi}_{\text{m},\vb*{k}}=P_m\psi_{\text{m},-\vb*{k}}$. Thus, the effective magnon Hamiltonian preserves the inversion symmetry when $t_2=0$.

\nocite{*}
\bibliography{references}

@CONTROL{REVTEX41Control}

@CONTROL{apsrev41Control,author="00",editor="1",pages="1",title="0",year="0"}

@article{Owerre2016AInsulator,
    title = {{A first theoretical realization of honeycomb topological magnon insulator}},
    year = {2016},
    journal = {Journal of Physics: Condensed Matter},
    author = {Owerre, S A},
    number = {38},
    month = {9},
    pages = {386001},
    volume = {28},
    doi = {10.1088/0953-8984/28/38/386001},
    issn = {0953-8984}
}

@article{Kargarian2016AmpereanInsulators,
    title = {{Amperean Pairing at the Surface of Topological Insulators}},
    year = {2016},
    journal = {Physical Review Letters},
    author = {Kargarian, Mehdi and Efimkin, Dmitry K. and Galitski, Victor},
    number = {7},
    month = {8},
    pages = {076806},
    volume = {117},
    doi = {10.1103/PhysRevLett.117.076806},
    issn = {0031-9007}
}

@article{Xiao2010BerryProperties,
    title = {{Berry phase effects on electronic properties}},
    year = {2010},
    journal = {Reviews of Modern Physics},
    author = {Xiao, Di and Chang, Ming-Che and Niu, Qian},
    number = {3},
    month = {7},
    pages = {1959--2007},
    volume = {82},
    doi = {10.1103/RevModPhys.82.1959},
    issn = {0034-6861}
}

@article{GomezAlbarracin2021ChiralLattice,
    title = {{Chiral phase transition and thermal Hall effect in an anisotropic spin model on the kagome lattice}},
    year = {2021},
    journal = {Physical Review B},
    author = {G{\'{o}}mez Albarrac{\'{i}}n, F. A. and Rosales, H. D. and Pujol, P.},
    number = {5},
    month = {2},
    volume = {103},
    doi = {10.1103/PhysRevB.103.054405},
    issn = {2469-9950}
}

@article{Hasan2010Colloquium:Insulators,
    title = {{Colloquium: Topological insulators}},
    year = {2010},
    journal = {Reviews of Modern Physics},
    author = {Hasan, M. Z. and Kane, C. L.},
    number = {4},
    month = {11},
    pages = {3045--3067},
    volume = {82},
    doi = {10.1103/RevModPhys.82.3045},
    issn = {0034-6861}
}

@article{Xiao2012CoupledDichalcogenides,
    title = {{Coupled Spin and Valley Physics in Monolayers of <math display="inline"> <msub> <mi>MoS</mi> <mn>2</mn> </msub> </math> and Other Group-VI Dichalcogenides}},
    year = {2012},
    journal = {Physical Review Letters},
    author = {Xiao, Di and Liu, Gui-Bin and Feng, Wanxiang and Xu, Xiaodong and Yao, Wang},
    number = {19},
    month = {5},
    pages = {196802},
    volume = {108},
    doi = {10.1103/PhysRevLett.108.196802},
    issn = {0031-9007}
}

@article{Pershoguba2018DiracFerromagnets,
    title = {{Dirac Magnons in Honeycomb Ferromagnets}},
    year = {2018},
    journal = {Physical Review X},
    author = {Pershoguba, Sergey S. and Banerjee, Saikat. and Lashley, J. C. and Park, Jihwey and {\AA}gren, Hans and Aeppli, Gabriel and Balatsky, Alexander V.},
    number = {1},
    month = {1},
    pages = {011010},
    volume = {8},
    doi = {10.1103/PhysRevX.8.011010},
    issn = {2160-3308}
}

@article{Murakami2003DissipationlessTemperature,
    title = {{Dissipationless Quantum Spin Current at Room Temperature}},
    year = {2003},
    journal = {Science},
    author = {Murakami, Shuichi and Nagaosa, Naoto and Zhang, Shou-Cheng},
    number = {5638},
    month = {9},
    pages = {1348--1351},
    volume = {301},
    doi = {10.1126/science.1087128},
    issn = {0036-8075}
}

@article{Ideue2012EffectInsulators,
    title = {{Effect of lattice geometry on magnon Hall effect in ferromagnetic insulators}},
    year = {2012},
    journal = {Physical Review B},
    author = {Ideue, T. and Onose, Y. and Katsura, H. and Shiomi, Y. and Ishiwata, S. and Nagaosa, N. and Tokura, Y.},
    number = {13},
    month = {4},
    pages = {134411},
    volume = {85},
    doi = {10.1103/PhysRevB.85.134411},
    issn = {1098-0121}
}

@article{Zhao2017EnhancedField,
    title = {{Enhanced valley splitting in monolayer WSe2 due to magnetic exchange field}},
    year = {2017},
    journal = {Nature Nanotechnology},
    author = {Zhao, Chuan and Norden, Tenzin and Zhang, Peiyao and Zhao, Puqin and Cheng, Yingchun and Sun, Fan and Parry, James P. and Taheri, Payam and Wang, Jieqiong and Yang, Yihang and Scrace, Thomas and Kang, Kaifei and Yang, Sen and Miao, Guo-xing and Sabirianov, Renat and Kioseoglou, George and Huang, Wei and Petrou, Athos and Zeng, Hao},
    number = {8},
    month = {8},
    pages = {757--762},
    volume = {12},
    doi = {10.1038/nnano.2017.68},
    issn = {1748-3387}
}

@article{Holstein1940FieldFerromagnet,
    title = {{Field Dependence of the Intrinsic Domain Magnetization of a Ferromagnet}},
    year = {1940},
    journal = {Physical Review},
    author = {Holstein, T. and Primakoff, H.},
    number = {12},
    month = {12},
    pages = {1098--1113},
    volume = {58},
    doi = {10.1103/PhysRev.58.1098},
    issn = {0031-899X}
}

@article{Rosales2019FromInsulators,
    title = {{From frustrated magnetism to spontaneous Chern insulators}},
    year = {2019},
    journal = {Physical Review B},
    author = {Rosales, H. D. and Albarrac{\'{i}}n, F. A. Gómez and Pujol, P.},
    number = {3},
    month = {1},
    pages = {035163},
    volume = {99},
    doi = {10.1103/PhysRevB.99.035163},
    issn = {2469-9950}
}

@article{Schneeloch2022GaplessCrCl3,
    title = {{Gapless Dirac magnons in CrCl3}},
    year = {2022},
    journal = {npj Quantum Materials},
    author = {Schneeloch, John A. and Tao, Yu and Cheng, Yongqiang and Daemen, Luke and Xu, Guangyong and Zhang, Qiang and Louca, Despina},
    number = {1},
    month = {6},
    pages = {66},
    volume = {7},
    doi = {10.1038/s41535-022-00473-3},
    issn = {2397-4648}
}

@article{Li2022GiantFerromagnetism,
    title = {{Giant valley splitting in a MoTe2/MnSe2 van der Waals heterostructure with room-temperature ferromagnetism}},
    year = {2022},
    journal = {Materials Advances},
    author = {Li, Qianze and Zhang, Cai-xin and Wang, Dan and Chen, Ke-Qiu and Tang, Li-Ming},
    number = {6},
    pages = {2927--2933},
    volume = {3},
    doi = {10.1039/D1MA01196K},
    issn = {2633-5409}
}

@article{Norden2019GiantEffect,
    title = {{Giant valley splitting in monolayer WS2 by magnetic proximity effect}},
    year = {2019},
    journal = {Nature Communications},
    author = {Norden, Tenzin and Zhao, Chuan and Zhang, Peiyao and Sabirianov, Renat and Petrou, Athos and Zeng, Hao},
    number = {1},
    month = {9},
    pages = {4163},
    volume = {10},
    doi = {10.1038/s41467-019-11966-4},
    issn = {2041-1723}
}

@article{Andrei2020GrapheneTwist,
    title = {{Graphene bilayers with a twist}},
    year = {2020},
    journal = {Nature Materials},
    author = {Andrei, Eva Y. and MacDonald, Allan H.},
    number = {12},
    month = {12},
    pages = {1265--1275},
    volume = {19},
    doi = {10.1038/s41563-020-00840-0},
    issn = {1476-1122}
}

@article{Mook2021Interaction-StabilizedFerromagnets,
    title = {{Interaction-Stabilized Topological Magnon Insulator in Ferromagnets}},
    year = {2021},
    journal = {Physical Review X},
    author = {Mook, Alexander and Plekhanov, Kirill and Klinovaja, Jelena and Loss, Daniel},
    number = {2},
    month = {6},
    pages = {021061},
    volume = {11},
    doi = {10.1103/PhysRevX.11.021061},
    issn = {2160-3308}
}

@article{Owerre2017MagnonInteraction,
    title = {{Magnon Hall effect without Dzyaloshinskii–Moriya interaction}},
    year = {2017},
    journal = {Journal of Physics: Condensed Matter},
    author = {Owerre, S A},
    number = {3},
    month = {1},
    pages = {03LT01},
    volume = {29},
    doi = {10.1088/0953-8984/29/3/03LT01},
    issn = {0953-8984}
}

@article{Zyuzin2016MagnonAntiferromagnets,
    title = {{Magnon Spin Nernst Effect in Antiferromagnets}},
    year = {2016},
    journal = {Physical Review Letters},
    author = {Zyuzin, Vladimir A. and Kovalev, Alexey A.},
    number = {21},
    month = {11},
    pages = {217203},
    volume = {117},
    doi = {10.1103/PhysRevLett.117.217203},
    issn = {0031-9007}
}

@article{Chumak2015MagnonSpintronics,
    title = {{Magnon spintronics}},
    year = {2015},
    journal = {Nature Physics},
    author = {Chumak, A. V. and Vasyuchka, V. I. and Serga, A. A. and Hillebrands, B.},
    number = {6},
    month = {6},
    pages = {453--461},
    volume = {11},
    doi = {10.1038/nphys3347},
    issn = {1745-2473}
}

@article{Laurell2018MagnonInteractions,
    title = {{Magnon thermal Hall effect in kagome antiferromagnets with Dzyaloshinskii-Moriya interactions}},
    year = {2018},
    journal = {Physical Review B},
    author = {Laurell, Pontus and Fiete, Gregory A.},
    number = {9},
    month = {9},
    pages = {094419},
    volume = {98},
    doi = {10.1103/PhysRevB.98.094419},
    issn = {2469-9950}
}

@article{Kim2019MagnonAntiferromagnet,
    title = {{Magnon topology and thermal Hall effect in trimerized triangular lattice antiferromagnet}},
    year = {2019},
    journal = {Physical Review B},
    author = {Kim, Kyung-Su and Lee, Ki Hoon and Chung, Suk Bum and Park, Je-Geun},
    number = {6},
    month = {8},
    pages = {064412},
    volume = {100},
    doi = {10.1103/PhysRevB.100.064412},
    issn = {2469-9950}
}

@article{VinasBostrom2024Magnon-mediatedWire,
    title = {{Magnon-mediated topological superconductivity in a quantum wire}},
    year = {2024},
    journal = {Physical Review Research},
    author = {Vi{\~{n}}as Bostr{\"{o}}m, Florinda and Vi{\~{n}}as Bostr{\"{o}}m, Emil},
    number = {2},
    month = {5},
    pages = {L022042},
    volume = {6},
    doi = {10.1103/PhysRevResearch.6.L022042},
    issn = {2643-1564}
}

@article{Mland2024Many-bodyAltermagnets,
    title = {{Many-body effects on superconductivity mediated by double-magnon processes in altermagnets}},
    year = {2024},
    journal = {Physical Review B},
    author = {M{\ae}land, Kristian and Brekke, Bjørnulf and Sudb{\o}, Asle},
    number = {13},
    month = {4},
    pages = {134515},
    volume = {109},
    doi = {10.1103/PhysRevB.109.134515},
    issn = {2469-9950}
}

@article{Chen2022MasslessCrCl3,
    title = {{Massless Dirac magnons in the two dimensional van der Waals honeycomb magnet CrCl3}},
    year = {2022},
    journal = {2D Materials},
    author = {Chen, Lebing and Stone, Matthew B and Kolesnikov, Alexander I and Winn, Barry and Shon, Wonhyuk and Dai, Pengcheng and Chung, Jae-Ho},
    number = {1},
    month = {1},
    pages = {015006},
    volume = {9},
    doi = {10.1088/2053-1583/ac2e7a},
    issn = {2053-1583}
}

@article{Lu2020Meron-likeCrCl3,
    title = {{Meron-like topological spin defects in monolayer CrCl3}},
    year = {2020},
    journal = {Nature Communications},
    author = {Lu, Xiaobo and Fei, Ruixiang and Zhu, Linghan and Yang, Li},
    number = {1},
    month = {9},
    pages = {4724},
    volume = {11},
    doi = {10.1038/s41467-020-18573-8},
    issn = {2041-1723}
}

@article{Haldane1988ModelAnomaly,
    title = {{Model for a Quantum Hall Effect without Landau Levels: Condensed-Matter Realization of the "Parity Anomaly"}},
    year = {1988},
    journal = {Physical Review Letters},
    author = {Haldane, F. D. M.},
    number = {18},
    month = {10},
    pages = {2015--2018},
    volume = {61},
    doi = {10.1103/PhysRevLett.61.2015},
    issn = {0031-9007}
}

@article{Klitzing1980NewResistance,
    title = {{New Method for High-Accuracy Determination of the Fine-Structure Constant Based on Quantized Hall Resistance}},
    year = {1980},
    journal = {Physical Review Letters},
    author = {Klitzing, K. v. and Dorda, G. and Pepper, M.},
    number = {6},
    month = {8},
    pages = {494--497},
    volume = {45},
    doi = {10.1103/PhysRevLett.45.494},
    issn = {0031-9007}
}

@article{Kondo2022NonlinearCurrent,
    title = {{Nonlinear magnon spin Nernst effect in antiferromagnets and strain-tunable pure spin current}},
    year = {2022},
    journal = {Physical Review Research},
    author = {Kondo, Hiroki and Akagi, Yutaka},
    number = {1},
    month = {3},
    pages = {013186},
    volume = {4},
    doi = {10.1103/PhysRevResearch.4.013186},
    issn = {2643-1564}
}

@article{Onose2010ObservationEffect,
    title = {{Observation of the Magnon Hall Effect}},
    year = {2010},
    journal = {Science},
    author = {Onose, Y. and Ideue, T. and Katsura, H. and Shiomi, Y. and Nagaosa, N. and Tokura, Y.},
    number = {5989},
    month = {7},
    pages = {297--299},
    volume = {329},
    doi = {10.1126/science.1188260},
    issn = {0036-8075}
}

@article{Zhang2026ObservationStates,
    title = {{Observation of Topological Magnon Edge States}},
    year = {2026},
    journal = {ACS Nano},
    author = {Zhang, Jihai and Zhang, Meng-Han and Li, Peigen and Liu, Zizhao and Tao, Ye and Wang, Hongkun and Yao, Dao-Xin and Guo, Donghui and Zhong, Dingyong},
    number = {1},
    month = {1},
    pages = {700--708},
    volume = {20},
    doi = {10.1021/acsnano.5c14794},
    issn = {1936-0851}
}

@article{Zollner2019ProximityDependence,
    title = {{Proximity exchange effects in MoSe2 and WSe2 heterostructures CrI3: Twist angle, layer, and gate dependence}},
    year = {2019},
    journal = {Physical Review B},
    author = {Zollner, Klaus and Faria Junior, Paulo E. and Fabian, Jaroslav},
    number = {8},
    month = {8},
    pages = {085128},
    volume = {100},
    doi = {10.1103/PhysRevB.100.085128},
    issn = {2469-9950}
}

@article{Onoda2003QuantizedMetals,
    title = {{Quantized Anomalous Hall Effect in Two-Dimensional Ferromagnets: Quantum Hall Effect in Metals}},
    year = {2003},
    journal = {Physical Review Letters},
    author = {Onoda, Masaru and Nagaosa, Naoto},
    number = {20},
    month = {5},
    pages = {206601},
    volume = {90},
    doi = {10.1103/PhysRevLett.90.206601},
    issn = {0031-9007}
}

@article{Thouless1982QuantizedPotential,
    title = {{Quantized Hall Conductance in a Two-Dimensional Periodic Potential}},
    year = {1982},
    journal = {Physical Review Letters},
    author = {Thouless, D. J. and Kohmoto, M. and Nightingale, M. P. and den Nijs, M.},
    number = {6},
    month = {8},
    pages = {405--408},
    volume = {49},
    doi = {10.1103/PhysRevLett.49.405},
    issn = {0031-9007}
}

@article{Qiao2010QuantumEffects,
    title = {{Quantum anomalous Hall effect in graphene from Rashba and exchange effects}},
    year = {2010},
    journal = {Physical Review B},
    author = {Qiao, Zhenhua and Yang, Shengyuan A. and Feng, Wanxiang and Tse, Wang-Kong and Ding, Jun and Yao, Yugui and Wang, Jian and Niu, Qian},
    number = {16},
    month = {10},
    pages = {161414},
    volume = {82},
    doi = {10.1103/PhysRevB.82.161414},
    issn = {1098-0121}
}

@article{Bernevig2006QuantumWells,
    title = {{Quantum Spin Hall Effect and Topological Phase Transition in HgTe Quantum Wells}},
    year = {2006},
    journal = {Science},
    author = {Bernevig, B. Andrei and Hughes, Taylor L. and Zhang, Shou-Cheng},
    number = {5806},
    month = {12},
    pages = {1757--1761},
    volume = {314},
    doi = {10.1126/science.1133734},
    issn = {0036-8075}
}

@article{Kane2005QuantumGraphene,
    title = {{Quantum Spin Hall Effect in Graphene}},
    year = {2005},
    journal = {Physical Review Letters},
    author = {Kane, C. L. and Mele, E. J.},
    number = {22},
    month = {11},
    pages = {226801},
    volume = {95},
    doi = {10.1103/PhysRevLett.95.226801},
    issn = {0031-9007}
}

@article{Yelon1971RenormalizationEffects,
    title = {{Renormalization of Large-Wave-Vector Magnons in Ferromagnetic CrBr3 Studied by Inelastic Neutron Scattering: Spin-Wave Correlation Effects}},
    year = {1971},
    journal = {Physical Review B},
    author = {Yelon, W. B. and Silberglitt, Richard},
    number = {7},
    month = {10},
    pages = {2280--2286},
    volume = {4},
    doi = {10.1103/PhysRevB.4.2280},
    issn = {0556-2805}
}

@article{Matsumoto2011RotationalEffect,
    title = {{Rotational motion of magnons and the thermal Hall effect}},
    year = {2011},
    journal = {Physical Review B},
    author = {Matsumoto, Ryo and Murakami, Shuichi},
    number = {18},
    month = {11},
    pages = {184406},
    volume = {84},
    doi = {10.1103/PhysRevB.84.184406},
    issn = {1098-0121}
}

@article{Cheng2016SpinAntiferromagnets,
    title = {{Spin Nernst Effect of Magnons in Collinear Antiferromagnets}},
    year = {2016},
    journal = {Physical Review Letters},
    author = {Cheng, Ran and Okamoto, Satoshi and Xiao, Di},
    number = {21},
    month = {11},
    pages = {217202},
    volume = {117},
    doi = {10.1103/PhysRevLett.117.217202},
    issn = {0031-9007}
}

@article{Zhang2018StrongHeterostructures,
    title = {{Strong magnetization and Chern insulators in compressed graphene/CrI3 van der Waals heterostructures}},
    year = {2018},
    journal = {Physical Review B},
    author = {Zhang, Jiayong and Zhao, Bao and Zhou, Tong and Xue, Yang and Ma, Chunlan and Yang, Zhongqin},
    number = {8},
    month = {2},
    pages = {085401},
    volume = {97},
    doi = {10.1103/PhysRevB.97.085401},
    issn = {2469-9950}
}

@article{Rohling2018SuperconductivityMagnons,
    title = {{Superconductivity induced by interfacial coupling to magnons}},
    year = {2018},
    journal = {Physical Review B},
    author = {Rohling, Niklas and Fj{\ae}rbu, Eirik Løhaugen and Brataas, Arne},
    number = {11},
    month = {3},
    pages = {115401},
    volume = {97},
    doi = {10.1103/PhysRevB.97.115401},
    issn = {2469-9950}
}

@article{Matsumoto2011TheoreticalFerromagnets,
    title = {{Theoretical Prediction of a Rotating Magnon Wave Packet in Ferromagnets}},
    year = {2011},
    journal = {Physical Review Letters},
    author = {Matsumoto, Ryo and Murakami, Shuichi},
    number = {19},
    month = {5},
    pages = {197202},
    volume = {106},
    doi = {10.1103/PhysRevLett.106.197202},
    issn = {0031-9007}
}

@article{Katsura2010TheoryMagnets,
    title = {{Theory of the Thermal Hall Effect in Quantum Magnets}},
    year = {2010},
    journal = {Physical Review Letters},
    author = {Katsura, Hosho and Nagaosa, Naoto and Lee, Patrick A.},
    number = {6},
    month = {2},
    pages = {066403},
    volume = {104},
    doi = {10.1103/PhysRevLett.104.066403},
    issn = {0031-9007}
}

@article{Kawano2019ThermalAntiferromagnet,
    title = {{Thermal Hall effect and topological edge states in a square-lattice antiferromagnet}},
    year = {2019},
    journal = {Physical Review B},
    author = {Kawano, Masataka and Hotta, Chisa},
    number = {5},
    month = {2},
    pages = {054422},
    volume = {99},
    doi = {10.1103/PhysRevB.99.054422},
    issn = {2469-9950}
}

@article{Mook2019ThermalAntiferromagnets,
    title = {{Thermal Hall effect in noncollinear coplanar insulating antiferromagnets}},
    year = {2019},
    journal = {Physical Review B},
    author = {Mook, Alexander and Henk, Jürgen and Mertig, Ingrid},
    number = {1},
    month = {1},
    pages = {014427},
    volume = {99},
    doi = {10.1103/PhysRevB.99.014427},
    issn = {2469-9950}
}

@article{Chatzichrysafis2025ThermalScattering,
    title = {{Thermal Hall effect of magnons from many-body skew scattering}},
    year = {2025},
    journal = {Physical Review B},
    author = {Chatzichrysafis, Dimos and Mook, Alexander},
    number = {13},
    month = {4},
    pages = {134405},
    volume = {111},
    doi = {10.1103/PhysRevB.111.134405},
    issn = {2469-9950}
}

@article{Matsumoto2014ThermalInteraction,
    title = {{Thermal Hall effect of magnons in magnets with dipolar interaction}},
    year = {2014},
    journal = {Physical Review B},
    author = {Matsumoto, Ryo and Shindou, Ryuichi and Murakami, Shuichi},
    number = {5},
    month = {2},
    pages = {054420},
    volume = {89},
    doi = {10.1103/PhysRevB.89.054420},
    issn = {1098-0121}
}

@article{Hirschberger2015ThermalMagnet,
    title = {{Thermal Hall Effect of Spin Excitations in a Kagome Magnet}},
    year = {2015},
    journal = {Physical Review Letters},
    author = {Hirschberger, Max and Chisnell, Robin and Lee, Young S. and Ong, N. P.},
    number = {10},
    month = {9},
    pages = {106603},
    volume = {115},
    doi = {10.1103/PhysRevLett.115.106603},
    issn = {0031-9007}
}

@article{Park2020ThermalInteraction,
    title = {{Thermal Hall Effect, Spin Nernst Effect, and Spin Density Induced by a Thermal Gradient in Collinear Ferrimagnets from Magnon-Phonon Interaction}},
    year = {2020},
    journal = {Nano Letters},
    author = {Park, Sungjoon and Nagaosa, Naoto and Yang, Bohm Jung},
    number = {4},
    volume = {20},
    doi = {10.1021/acs.nanolett.0c00363},
    issn = {15306992}
}

@article{Fujiwara2022ThermalSystems,
    title = {{Thermal Hall responses in frustrated honeycomb spin systems}},
    year = {2022},
    journal = {Physical Review B},
    author = {Fujiwara, Kosuke and Kitamura, Sota and Morimoto, Takahiro},
    number = {3},
    month = {7},
    pages = {035113},
    volume = {106},
    doi = {10.1103/PhysRevB.106.035113},
    issn = {2469-9950}
}

@article{Liu2013Three-bandDichalcogenides,
    title = {{Three-band tight-binding model for monolayers of group-VIB transition metal dichalcogenides}},
    year = {2013},
    journal = {Physical Review B},
    author = {Liu, Gui-Bin and Shan, Wen-Yu and Yao, Yugui and Yao, Wang and Xiao, Di},
    number = {8},
    month = {8},
    pages = {085433},
    volume = {88},
    doi = {10.1103/PhysRevB.88.085433},
    issn = {1098-0121}
}

@article{Qi2011TopologicalSuperconductors,
    title = {{Topological insulators and superconductors}},
    year = {2011},
    journal = {Reviews of Modern Physics},
    author = {Qi, Xiao-Liang and Zhang, Shou-Cheng},
    number = {4},
    month = {10},
    pages = {1057--1110},
    volume = {83},
    doi = {10.1103/RevModPhys.83.1057},
    issn = {0034-6861}
}

@article{Park2019TopologicalAntiferromagnets,
    title = {{Topological magnetoelastic excitations in noncollinear antiferromagnets}},
    year = {2019},
    journal = {Physical Review B},
    author = {Park, Sungjoon and Yang, Bohm Jung},
    number = {17},
    volume = {99},
    doi = {10.1103/PhysRevB.99.174435},
    issn = {24699969}
}

@article{Zhang2020TopologicalMagnet,
    title = {{Topological magnon bands in a room-temperature kagome magnet}},
    year = {2020},
    journal = {Physical Review B},
    author = {Zhang, H. and Feng, X. and Heitmann, T. and Kolesnikov, A. I. and Stone, M. B. and Lu, Y.-M. and Ke, X.},
    number = {10},
    month = {3},
    pages = {100405},
    volume = {101},
    doi = {10.1103/PhysRevB.101.100405},
    issn = {2469-9950}
}

@article{Go2019TopologicalNumbers,
    title = {{Topological Magnon-Phonon Hybrid Excitations in Two-Dimensional Ferromagnets with Tunable Chern Numbers}},
    year = {2019},
    journal = {Physical Review Letters},
    author = {Go, Gyungchoon and Kim, Se Kwon and Lee, Kyung-Jin},
    number = {23},
    month = {12},
    pages = {237207},
    volume = {123},
    doi = {10.1103/PhysRevLett.123.237207},
    issn = {0031-9007}
}

@article{Zhang2010TopologicalEffect,
    title = {{Topological Nature of the Phonon Hall Effect}},
    year = {2010},
    journal = {Physical Review Letters},
    author = {Zhang, Lifa and Ren, Jie and Wang, Jian-Sheng and Li, Baowen},
    number = {22},
    month = {11},
    pages = {225901},
    volume = {105},
    doi = {10.1103/PhysRevLett.105.225901},
    issn = {0031-9007}
}

@article{Fujiwara2025TopologicalInteractions,
    title = {{Topological phases of electrons induced by electron-magnon interactions}},
    year = {2025},
    journal = {Physical Review B},
    author = {Fujiwara, Kosuke and Morimoto, Takahiro},
    number = {23},
    month = {12},
    pages = {235128},
    volume = {112},
    doi = {10.1103/bx1x-qc78},
    issn = {2469-9950}
}

@article{Chen2018TopologicalCrI3,
    title = {{Topological Spin Excitations in Honeycomb Ferromagnet CrI3}},
    year = {2018},
    journal = {Physical Review X},
    author = {Chen, Lebing and Chung, Jae-Ho and Gao, Bin and Chen, Tong and Stone, Matthew B. and Kolesnikov, Alexander I. and Huang, Qingzhen and Dai, Pengcheng},
    number = {4},
    month = {11},
    pages = {041028},
    volume = {8},
    doi = {10.1103/PhysRevX.8.041028},
    issn = {2160-3308}
}

@article{Mland2023TopologicalStates,
    title = {{Topological superconductivity mediated by magnons of helical magnetic states}},
    year = {2023},
    journal = {Physical Review B},
    author = {M{\ae}land, Kristian and Abnar, Sara and Benestad, Jacob and Sudb{\o}, Asle},
    number = {22},
    month = {12},
    pages = {224515},
    volume = {108},
    doi = {10.1103/PhysRevB.108.224515},
    issn = {2469-9950}
}

@article{Mland2023TopologicalMagnons,
    title = {{Topological Superconductivity Mediated by Skyrmionic Magnons}},
    year = {2023},
    journal = {Physical Review Letters},
    author = {M{\ae}land, Kristian and Sudb{\o}, Asle},
    number = {15},
    month = {4},
    pages = {156002},
    volume = {130},
    doi = {10.1103/PhysRevLett.130.156002},
    issn = {0031-9007}
}

@article{Owerre2017TopologicalAntiferromagnets,
    title = {{Topological thermal Hall effect in frustrated kagome antiferromagnets}},
    year = {2017},
    journal = {Physical Review B},
    author = {Owerre, S. A.},
    number = {1},
    month = {1},
    pages = {014422},
    volume = {95},
    doi = {10.1103/PhysRevB.95.014422},
    issn = {2469-9950}
}

@article{Gong2019Two-dimensionalDevices,
    title = {{Two-dimensional magnetic crystals and emergent heterostructure devices}},
    year = {2019},
    journal = {Science},
    author = {Gong, Cheng and Zhang, Xiang},
    number = {6428},
    month = {2},
    volume = {363},
    doi = {10.1126/science.aav4450},
    issn = {0036-8075}
}

@article{Ezawa2012Valley-PolarizedSilicene,
    title = {{Valley-Polarized Metals and Quantum Anomalous Hall Effect in Silicene}},
    year = {2012},
    journal = {Physical Review Letters},
    author = {Ezawa, Motohiko},
    number = {5},
    month = {8},
    pages = {055502},
    volume = {109},
    doi = {10.1103/PhysRevLett.109.055502},
    issn = {0031-9007}
}

@article{Zhong2017VanValleytronics,
    title = {{Van der Waals engineering of ferromagnetic semiconductor heterostructures for spin and valleytronics}},
    year = {2017},
    journal = {Science Advances},
    author = {Zhong, Ding and Seyler, Kyle L. and Linpeng, Xiayu and Cheng, Ran and Sivadas, Nikhil and Huang, Bevin and Schmidgall, Emma and Taniguchi, Takashi and Watanabe, Kenji and McGuire, Michael A. and Yao, Wang and Xiao, Di and Fu, Kai-Mei C. and Xu, Xiaodong},
    number = {5},
    month = {5},
    volume = {3},
    doi = {10.1126/sciadv.1603113},
    issn = {2375-2548}
}

\end{document}